\documentclass[12pt]{amsart}
\usepackage{amsmath}
\usepackage{epsfig}  		% For postscript
\usepackage{epic,eepic}       % For epic and eepic output from xfig
\usepackage{hyperref}
\hypersetup{
    colorlinks=true,
    linkcolor=blue,
    filecolor=magenta,
    urlcolor=cyan,
}%%
\usepackage{mathrsfs}
\usepackage{bbm}

\theoremstyle{plain}
\newtheorem{theorem}{Theorem}[section]

\newtheorem{proposition}[theorem]{Proposition}
\newtheorem{lemma}[theorem]{Lemma}

\newtheorem{claim}[theorem]{Claim}
\newtheorem{definition}[theorem]{Definition}

\newtheorem{example}[theorem]{Example}
\numberwithin{equation}{section}

\newcommand{\R}{\mathbf{R}}
\def\1{{1\hskip-2.5pt{\rm l}}}

\def\SS{\mathcal{S}}

\def\a{\alpha}
\def\b{\beta}
\def\lm{\lambda}
\def\Lm{\Lambda}
\def\eps{\varepsilon}
\def\s{\sigma}

\def\d{\delta}

\def\lm{\lambda}

\def\g{\gamma}
\def\H2{\widehat{\mathcal{H}}^2}

\def\G8{\norm{G}_{\infty}}
\def\conv{\rm conv}

\def\R{\mathbb{R}}
\def\Y{\mathcal{Y}}
\def\N{\mathbb{N}}
\def\C{\textbf{\textrm{C}}}
\def\l{\ell}

\def\G{\Gamma}

\def\vd{\underline{v}}
\def\vu{\bar{v}}
\def\Pr{\mathbb{P}}

\def\cl{\text{cl}\,}

\def\E{\mathbb{E}}

\def\eps{\varepsilon}
\def\Rr{\mathcal{R}}

\def\S{\Sigma}
\def\C{c}

\def\cv{\text{Cav}\, u}
\def\Cv{\emph{Cav}\, u}
\def\ccv{(\text{Cav}\, u)(\pi_M)}
\def\Ccv{(\emph{Cav}\, u)(\pi_M)}
\def\cavu{(\text{Cav}\, u)}
\def\conv{\text{conv}}

\def\Graph{\text{Graph}}
\def\p{\pi_M}
\def\dk{\Delta(K)}
\usepackage{graphicx}

\newcommand{\ignore}[1]{}

\begin{document}

%%
%% The title of the paper goes here.  Edit to your title.
%%

\title{Markovian Persuasion}

%%
%% Now edit the following to give your name and address:
%%

\author{Ehud Lehrer}
\address{%School of Mathematical Sciences,
Tel Aviv University, Tel Aviv 69978,
Israel. }
\email{lehrer@post.tau.ac.il}
\thanks{\noindent The first author acknowledges the support of  grants ISF 591/21 and DFG KA 5609/1-1.}

%%\urladdr{www.math.sc.edu/$\sim$howard} % Delete if not wanted.

\author{Dimitry Shaiderman}
\address{%School of Mathematical Sciences,
Tel Aviv University, Tel Aviv 69978,
Israel.}
\email{dima.shaiderman@gmail.com}

\maketitle
\thispagestyle{empty}

%%
%% If there are three of more authors they are added in the obvious
%% way.
%%

%%%
%%% The following is for the abstract.  The abstract is optional and
%%% if not used just delete, or comment out, the following.
%%%

\begin{abstract}

In the classical Bayesian persuasion model an informed player and an uninformed one engage in a static interaction. The informed player, the sender, knows the state of nature, while  the uninformed one, the receiver, does not. The informed player partially shares his private information with the receiver and the latter then, based on her belief about the state, takes an action. This action determines, together with the state of nature, the utility of both players.  We consider  a dynamic Bayesian persuasion situation where the state of nature evolves according to a Markovian law. In this repeated persuasion model an optimal disclosure strategy of the sender should, at any period, balance between getting high stage payoff and future implications on the receivers' beliefs. We discuss optimal strategies under different discount factors and characterize when the asymptotic value achieves the maximal value possible.
\end{abstract}

\bigskip
\noindent \textbf{Date:} \today

%%
%%  LaTeX will not make the title for the paper unless told to do so.
%%  This is done by uncommenting the following.
%%

\newpage

\pagebreak  \setcounter{page}{1}
\normalsize
%%
%% LaTeX can automatically make a table of contents.  This is done by
%% uncommenting the following:
%%

%%\tableofcontents

%%
%%  To enter text is easy.  Just type it.  A blank line starts a new
%%  paragraph.
%%

%%%%%%%%%%%%%%%%%%%
\section{Introduction}
%%%%%%%%%%%%%%%%%%%

The literature devoted to Bayesian persuasion studies optimal policies by which an informed agent, the sender,  discloses information to an uninformed agent, the receiver.
Kamenika and Gentzkow \cite{Kamenika} present a case where a prosecutor, who is fully informed about the state of nature, i.e., whether the suspect is guilty or
innocent,  wishes to persuade a judge, e.g to convict the suspect of a crime. This is a static scenario: upon the prosecutor's disclosure, the judge takes a  decision and the game is over.

In this paper we study a dynamic model where the interaction between the sender, who has a commitment power, and the receiver evolves over time. At each period nature randomly chooses a state and the sender gets informed about it. The decision  of how
much information to disclose to the receiver at each stage is the choice of the sender.
The latter publicly announces an information provision policy, to which he is committed throughout.
The receiver knows the details of this policy and thus, when she receives a signal from the sender she updates her belief about the true state.
She then takes an action that is based solely on her posterior belief.
This action, together with the realized state, determines not only her own payoff, but also that of the sender. In short, in this dynamic signaling
interaction, the sender is in charge of the informational aspects of the interaction (talking), while the receiver is in charge of doing. Neither may affect the evolution of the realized states, which is entirely exogenous.

We assume that the evolution of states follows a Markov chain. The dynamic of the receiver's belief is governed by both the disclosure policy of the sender and a Markov transition matrix. After observing the signal sent by the sender, the receiver updates her belief according to Bayes' law. Due to the Markovian transition law, this belief shifts to another belief. Since all terms are known to both players, the nature of the shift from prior belief to a posterior one is also known to both.

When committing to an information provision policy, the sender takes into account the resulting posterior belief of the receiver, which has two effects. The first is on the action taken by the receiver and directly on his own stage payoff. In a dynamic setting, as discussed here, it has also an effect on the belief over future states, and consequently on future payoffs. The posterior resulting from a Bayesian updating in one period is then shifted by the Markov transition matrix and becomes the initial belief in the next one.
Balancing between these two, potentially contradicting effects, is the dynamic programming problem faced by the sender.

In this paper we  study the long-run optimal values that the sender may achieve in an irreducible and aperiodic Markov chain.\footnote{In a companion paper we examine, based on the current study, the general case.} This value depends on his discount factor. As in Aumann and Maschler \cite{Aumann} and in Kamenika and Gentzkow \cite{Kamenika}, the optimal signaling strategy follows a splitting scheme of beliefs in accordance with the concavification of a certain function\footnote{That is, the lowest concave function above a given function.}. While in a static model this function is the payoff function of the sender, here, this function takes account of the underlying dynamics and combines present and future payoffs.

The first set of results deal with a patient sender. It turns out that the stationary distribution of the Markov chain plays a central role. The reason is that in case the sender keeps mute, the receiver's beliefs become converge, very fast,\footnote{At an exponential rate (see e.g., Theorem 4.9 in \cite{Peres}).} to the stationary distribution.

We show that as the sender becomes increasingly patient,\footnote{The method of treating patient players in not new to the literature of dynamic interactions. Fudenberg and Maskin \cite{Fudenberg} introduced a Folk Theorem characterizing the set of equilibria in a repeated game played by patient players. Typically in these kind of games, it is impossible to provide a description of the value or of the set of equilibria that correspond to a specific discount factor more informative than that given by Bellman equation (see Abreu et.\ al.\ \cite{Abreu}).} the optimal values converges to a certain level, called the asymptotic value. Moreover, this value cannot exceed the optimal level obtained in the static case when the prior belief is the stationary distribution. A natural  question  is under what conditions this static optimal value can be obtained in the dynamic model. Our main theorem provides a full characterization. In order to describe it we introduce the notion of an absorbing set.

A set $C$ of prior distributions is said to be absorbing, if any  distribution $p$ in $C$  evolves by the Markov law to another distribution  which is within the convex boundaries of $C$. The stationary distribution, for instance, is absorbing as it shifts to itself. The main theorem roughly asserts that the asymptotic value is as high as it can get (i.e., the static value at the stationary distribution), precisely when there is an absorbing set by which the optimal static value is defined (as a weighted average). The intuition behind this result is the following. If any set $A$ by means of which the optimal static value is defined is non-absorbing, then regardless of the policy the sender uses, the posterior beliefs of the receiver would exit the boundaries of $A$ infinitely often, at times having positive limit frequency. Upon each such exit, the sender's payoff would take a hit which he would not be able to compensate for in subsequent time periods. Therefore, the dream scenario for a patient sender is to find a set $A$ by means of which the optimal static value is defined, so that he could confine the posteriors beliefs of the receiver within the convex boundaries of $A$ throughout the entire game.

The second type of results is non-asymptotic in nature. For a certain region around the stationary distribution we provide a closed form expression for the values corresponding to any level of patience. In the case where this region includes a neighborhood of the stationary distribution, we give asymptotically effective\footnote{That is, bounds whose difference is arbitrarily small for a patient enough sender.} two-sided bounds for the values  corresponding to any level of patience. Moreover, the effectiveness of those bounds depends on the geometry and size of the described neighborhood.

The closest paper to the present one is that of Renault et.\ al.\ \cite{Solan}. It deals
with a specific type of Markov chains, homothetic ones, and with a utility function which gives a fixed amount to the sender in a certain region and zero in its complement. Renault et.\ al.\ \cite{Solan} show that in this model the optimal information provision policy is a greedy one (starting from a certain random time period on). Namely, the sender's greedy policy instructs him to maximize his current stage payoff at any time period, ignoring future effects of this policy. Also closely related to our work are the works of Ely \cite{Ely} and Farhadi and  Teneketzis \cite{Farhadi},  which deal with a case where there are two states, whose evolution is described by a Markov chain with one absorbing state. In these two works, the receiver is interested in detecting the jump to the absorbing state, whereas the sender seeks to prolong the time to detection of such a jump.

In a broader context, our paper can be ascribed to a vast growing literature on dynamic information design problems, specifically with the analysis of situations in which the sender is required to convey signals sequentially to a receiver, which he may base on his additional private information. Without dwelling upon the exact details of their models, whose frameworks cover diverse methodological approaches, we refer the reader to Mailath and Samuelson \cite{Mailath}, Honryo \cite{Honryo}, Phelan \cite{Phelan}, Lingenbrink and Iyer \cite{Lingenbrink}, Wiseman \cite{Wiseman}, Athey and Bagwell \cite{Athey}, Arieli and Babichenko \cite{Arieli}, Arieli et.\ al.\ \cite{Arieli2}, Dziuda and Gradwohl \cite{Dziuda}, Escobar and Toikka \cite{Escobar}, Ganglmair and Tarantino \cite{Gang}, and Augenblick and Bodoh-Creed \cite{Aug} for an acquaintance with the literature.

The paper is organized as follows. Section \ref{sec: model} presents the model and an example by which we explain the notations, new concepts and the main results. The asymptotic results as well as the main theorem are provided in Section \ref{sec:Main Theorem}. Results related to non-asymptotic values are given in Section \ref{sec:non-asymptotic}. Section \ref{sec:homothety} provides a characterization of homothetic matrices in terms of the asymptotic value. The proofs are given in Section \ref{sec:proofs}.

\bigskip

%%%%%%%%%%%%%%%%%%%
\section{The Model}\label{sec: model}
%%%%%%%%%%%%%%%%%%%

Let $K=\{1,...,k\}$ be a finite set of states. Assume that $(X_n)_{n \geq 1}$ is an irreducible and aperiodic Markov chain over $K$ with prior probability $p \in \Delta(K)$ and a transition rule given by the stochastic matrix $M$. We assume that $(X_n)_{n \geq 1}$ are defined on the probability space $(\Omega, \mathcal{F},\Pr)$.

A \emph{sender} is an agent who is informed at each time period $n$ of the realized value $x_n$ of $X_n$. Upon obtaining this information a sender is prescribed to send a signal $s_n$, from a finite set of signals $S$ with cardinality at least $k$.\footnote{This assumption is in place to make sure the sender can disclose $(x_n)_n$.}

A \emph{receiver} is an agent who, at any time period $n$, is instructed to make a decision $b_n$ from a set of possible set of decisions $B$, assumed to be a compact metric space. This decision may take into account the first $n$ signals $s_1,...,s_n$ of the sender.

The payoffs of the sender and the receiver at time period $n$ are given by the utilities $v(x_n,b_n)$  and $w(x_n,b_n)$, respectively, so that they depend solely on the realized state $x_n$ and the decision $b_n$. Both the sender and the receiver discount their payoffs by a factor $\lm\in [0,1)$. We denote this game by $\G_{\lm}(p)$. As in the models of Renault et. al.\ \cite{Solan}, Ely \cite{Ely}, and Farhadi and  Teneketzis \cite{Farhadi} the receiver receives information only through the sender.
%does not observe his realized payoffs along the play of  $\G_{\lm}(p)$.

A \emph{signaling strategy} $\sigma$ of the sender in $\G_{\lm}(p)$ is described by a sequence of stage strategies $(\s_n)$, where each $\s_n$ is a mapping $\s_n : (K \times S)^{n-1}\times K \to \Delta(S).$ Thus, the signal $s_n$, sent by the sender at time $n$ is distributed by the lottery $\s_n$, which may  depend on all past states $x_1,...,x_{n-1}$ and past signals $s_1,...,s_n$ together with the current state $x_n$. Let $\S$ be the space of all signaling strategies.

A standard assumption in many Bayesian persuasion models is that of \emph{commitment} by the sender. That is, we assume that the sender commits to a signaling strategy $\s$ at the start of the game $\G_{\lm}(p)$, and makes it known to the receiver. The commitment assumption enables the receiver to update her beliefs on the distribution of states $(X_n)$ based on the signals $(s_n)$ he receives from the sender. Formally, by Kolmogorov's Extension Theorem, each signaling strategy $\s$ together with $(X_n)_{n \geq 1}$ induce a unique probability measure $\Pr_{p,\s}$ on the space $\Y = (K \times S)^{\N}$, determined by the laws
\begin{multline}\tag{1}
\Pr_{p,\s} (x_1,s_1,...,x_n,s_n) = \left(p(x_1)\prod_{i=1}^{n-1}M_{x_i,x_{i+1}}\right)\times \\ \left(\prod_{i=1}^n \s_i(x_1,s_1,...,x_{i-1},s_{i-1},x_i)(s_i)\right).
\end{multline}
Thus, the posterior probability $p_n^{\l}$ the receiver assigns to the event $\{X_n=\l\}$, given the signals $s_1,...,s_n$ and the strategy $\s$, is given by the formula
\begin{equation}\tag{2}
p_n^{\l} = \Pr_{p,\s}\left(X_n=\l\,|\, s_1,...,s_n\right).
\end{equation}
Set $p_n = (p_n^{\l})_{\l \in K}$. A second key assumption of our model is that the receiver's decision at any time period $n$ depends only on $p_n$. Such an assumption includes the natural situation in which the receiver seeks to maximize her expected payoff based on his current belief (e.g., Renault et.\ al.\ (2015)). Denote by $\theta : \Delta(K)\to B$ the decision policy of the receiver, that is, the mapping which depicts the decision of the receiver as a function of his belief. The last assumption of our model is that the function $u: \Delta(K) \to \R$ defined by $u(q)= \sum_{\l \in K} q^{\l} v(\l,\theta(q))$ is continuous. To summarize, our assumptions imply that the signaling strategy $\s$ of the sender determines his payoff at any time period $n$. Moreover, the total expected payoff to the sender in $\G_{\lm}(p)$ under the signaling strategy $\s$ can now be written as
\begin{equation}\tag{3}
\g_{\lm}(p,\s) := \E_{p,\s}\left[(1-\lm)\sum_{n=1}^{\infty} \lm^{n-1} u(p_n)\right],
\end{equation}
 where $\E_{p,\s}$ is the expectation w.r.t.\ $\Pr_{p,\s}$. The value of the game $\G_{\lm}(p)$ is $v_{\lm}(p) = \sup_{\s \in \S} \g_{\lm}(p,\s)$.
\begingroup
\renewcommand\thetheorem{1}
\begin{example}\label{ex:main}  Assume that $K$ consists of two states, $H$ and $L$.  Each probability measure $(p,1-p)$ over $K$ is determined by a parameter $p \in [0,1]$, corresponding to the distribution mass assigned to the state $H$ by the respective probability measure. Suppose that the receiver's set of decisions $B$ is $[0,1]$ and that when his belief is $(p,1-p)$ his decision policy $\theta$ chooses $p$.

As for the sender, when the decision made by the receiver is $p$ and the state is $H$, his utility is $2$ minus $3$ times the distance between $p$ and $\frac{1}{2}$. That is, $v(H,p)=2-3|p-\frac{1}{2}|$. In words, when $H$ is the realized state, the closer the decision of the receiver to $\frac{1}{2}$, the higher the payoff of the sender. However,
when the decision made by the receiver is $p$ and the state is $L$, his utility is $p/10$. That is, $v(L,p)=p/10$.  When $L$ is the realized state, the closer the decision of the receiver to $1$, the higher the payoff of the sender.

We obtain that $u((p,1-p))= pv(H,p)+(1-p)v(L,p)= p(2-3|p-\frac{1}{2}|)+(1-p)p/10$. The graph of $u$, which for the sake of convenience is plotted as a function of $p$, can be found on the left panel of Figure \emph{\ref{fugure:utility graph}}. Thus, $u$ is convex on the interval $[0,0.5]$ and concave on the interval $[0.5,1]$. This graph illustrates the short-term strategic incentives of the sender. Indeed, the minimal possible stage payoff, $0$,  occurs when the sender reveals that the realized state is $L$ \emph{(}corresponding to $p=0$\emph{)}. A revelation of the state $H$ \emph{(}corresponding to $p=1$\emph{)} would result in a payoff of $0.5$. The maximal payoff for the sender, $1.045$, is reached for $p=0.581$. As explained below \emph{(}see Example \emph{\ref{ex: example 2 continued}}\emph{)}, for that $p$ the optimal signaling strategy would instruct the sender to not reveal any information regarding the realized state.

In a dynamic model the amount of information revealed by the sender is a result of an interplay between the one-shot payoff, $u$, and the transition law governing the evolution of future states. The tension between these two factors is discussed in the rest of this paper.
\end{example}
\endgroup
\begingroup
\renewcommand\thetheorem{1}
\begin{figure}[htp] \centering{
\includegraphics[scale=0.33]{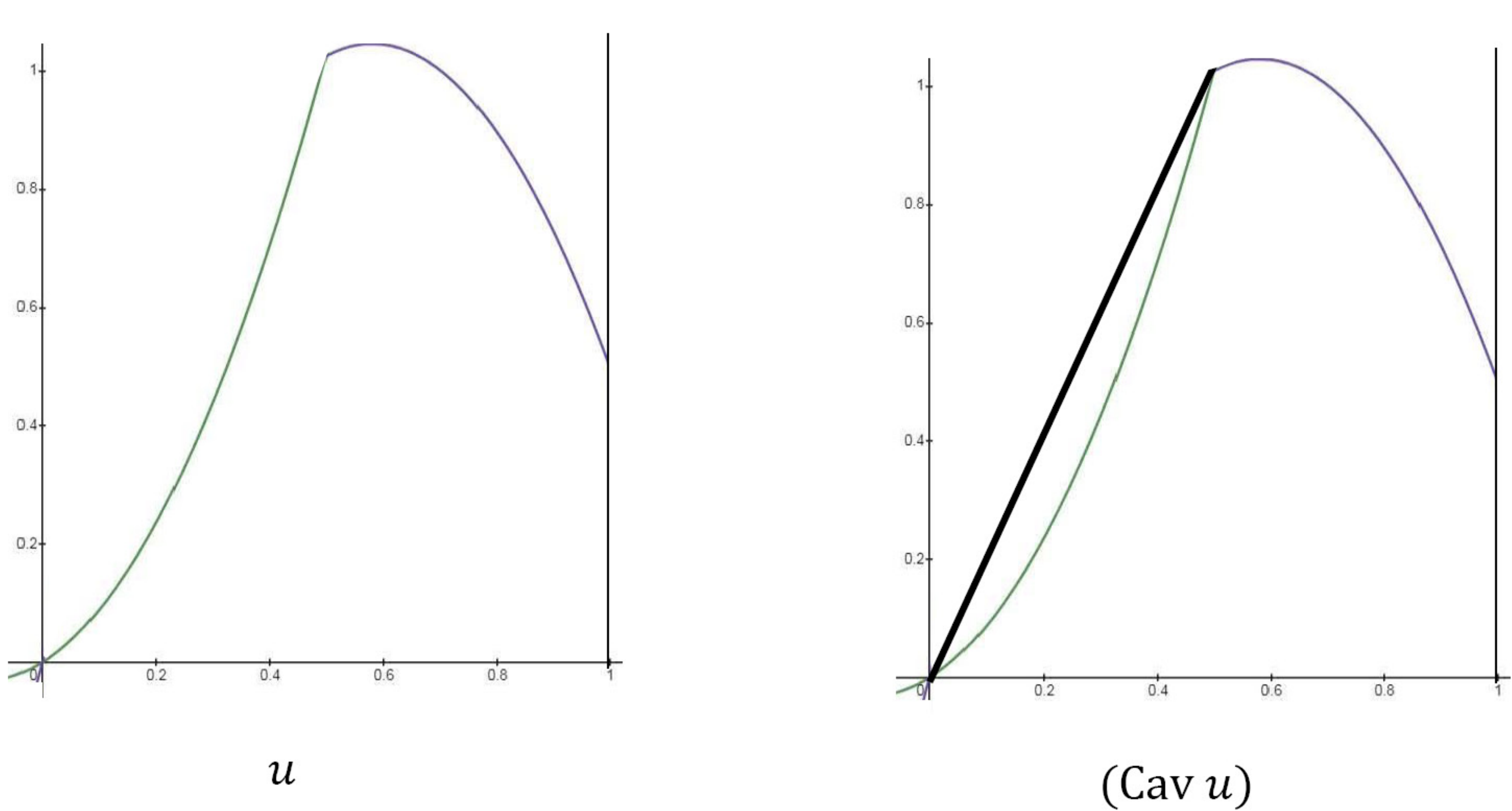}}
\caption{The graphs of $u$ and  ($\cv$)}
\label{fugure:utility graph}
\end{figure}
\endgroup
%%%%%%%%%%%%%%%%%%%

\bigskip

\section{The Main Theorem} \label{sec:Main Theorem}
%%%%%%%%%%%%%%%%%%%
\subsection{The Existence of Asymptotic Value.}
To state our first result we need to introduce some notations. First, let $\p$ be the unique stationary distribution of $M$. Second, for any function $g: \dk \to \R$, define the function  $(\text{Cav}\, g)$ by
\begin{equation*}
(\text{Cav}\, g) (q) := \inf\{h(q)\,:\, h: \dk \to \R\,\,  \text{concave},\, h \geq g \}, 	\	\	\ \ \forall q \in \dk.
\end{equation*}
To showcase the Cav operator in action, we show the graph of ($\cv$), where $u$ is that given in Example \ref{ex:main}, on the right-hand panel of Figure \ref{fugure:utility graph}.

Our first result reveals that the effect of $p \in \dk$ on the value $v_{\lm}(p)$  of a sufficiently patient sender, i.e., with $\lm$ close to $1$, is negligible compared to the effects of $u$ and $M$. Moreover, as the patience level $\lm$ gets closer to $1$ the sequences $v_{\lm}(p)$ converge for any $p$. Formally, this result is stated as follows.
\begingroup
\renewcommand\thetheorem{1}
\begin{theorem}\label{Thm1}
There exists a scalar $v_{\infty} \in \R$, $v_{\infty} \leq \Ccv$, such that for every $\eps>0$ and every $p \in \dk$ there exists $0<\delta<1$ such that
\begin{equation}\tag{4}
\vert v_{\lm}(p) - v_{\infty} \vert < \eps, 	\	\		\	\	\	 \forall \lm > \delta.
\end{equation}
\end{theorem}
\endgroup
As it turns out, the upper bound on $v_{\infty} $ described in Theorem \ref{Thm1} is tight. In our next section we shall give a geometric criteria on the achievement of this upper bound. To do so we begin by introducing and studying the notion of $M$-absorbing sets.

\subsection{$M$-absorbing sets.}
\begingroup
\renewcommand\thetheorem{1}
\begin{definition}
A non-empty set $C \subseteq  \Delta(K)$ is said to be $M$\emph{-absorbing} if\,\footnote{$\conv(C)$ is the convex hull of the set $C$.} $qM \in \emph{conv}(C)$ for every $q \in C$.
\end{definition}

\endgroup
The intuition  behind this choice of terminology is the following. Since $q \mapsto qM$ is a linear operator, if $C$ is $M$-absorbing, so in $\conv(C)$. However for $\conv(C)$, $M$-absorption exactly  describes the situation in which the image of $\conv(C)$ under $M$ lies inside (is absorbed in) $\conv(C)$. In a dual fashion, if $C \subseteq \dk$ is a closed convex $M$-absorbing set, then since by Krein-Milman Theorem\footnote{$\text{ext}(C)$ is the set of the extreme points of $C$: those points in $C$ that cannot be expressed as a convex combination of two distinct points in $C$.} $\conv(\text{ext}(C))=C$, we get that $\text{ext}(C)$ is also $M$-absorbing. This implies, in particular, that since $\dk$ is $M$-absorbing, so is the set of its extreme points (i.e., the set of all mass-point distributions). Lastly, note that since $\conv(C_1) \cup \conv(C_2) \subseteq \conv(C_1 \cup C_2)$, if $C_1$ and $C_2$ are both $M$-absorbing, then so is $C_1 \cup C_2$.

\vskip .5cm
%\begingroup
\renewcommand\thetheorem{2}
\begin{example} \label{ex:main-2}
\emph{[Example \ref{ex:main}, continued]} Suppose that the transition rule underlying  the interaction in Example \emph{\ref{ex:main}} is given by
\begin{equation}\label{eq: matrix}\tag{5}
  M=\begin{pmatrix}
  \ 0.1& 0.9\\
 0.6 & 0.4
\end{pmatrix}.
\end{equation}
Note that the stationary distribution $\p$ of $M$ equals $ (0.4,0.6)$ \emph{(}which corresponds to $ p=0.4$ in Figure \emph{\ref{fugure:utility graph})} and the pair $((0.4,0.6), (\emph{Cav}\, u) ((0.4,0.6))$ is on the straight segment of the graph of $(\emph{Cav}\, u)$ (see  the right panel of Figure \emph{\ref{fugure:utility graph}}).

Moreover, the set containing the points $(0,1)$ and $ (0.5,0.5)$ \emph{(}presented in Figure \emph{\ref{fugure:utility graph}} as the points $0$ and $0.5$\emph{)} is not $M$-absorbing. The reason is that $(0,1)$ is mapped by $M$ to $(0,1)M=(0.6 , 0.4)$, which is not in the convex hull of $(0,1)$ and $ (0.5,0.5)$.
\end{example}
%\endgroup

\vskip .5cm

In order to enhance the intuition about $M$-absorbing sets, assume that at any point $q \in \dk$ a football is passed in a straight line to the point $qM$. The orbit generated by the football at $q$ is the union of line segments $\bigcup_{n \geq 1}[qM^{n-1},qM^n]$, where $[x,y] = \{\a x +(1-\a)y:\, 0\leq \a \leq 1 \}$. When the set $C$ is $M$-absorbing, the orbit generated by the football whose starting point is $q \in C$ never exists of $\conv(C)$.

We proceed with some basic examples of $M$-absorbing sets. The simplest are the singleton $\{\p\}$ and the entire set, $\dk$. To describe additional examples, consider the $\l_1$-,$\l_2$-, and $\l_{\infty}$-norms on $\dk$, denoted by $\Vert q \Vert_{1} := \sum_{\l \in K}\vert q^{\l}\vert$, $\Vert q \Vert_{2} := \sqrt{\sum_{\l \in K}(q^{\l})^2}$ and $\Vert q \Vert_{\infty} := \max_{\l \in K} \vert q^{\l} \vert$, respectively, for  $q \in \dk$. Denote by $\Vert M\Vert_i$ the operator norm\footnote{ $\Vert M\Vert_i=\max_{\Vert x\Vert_i = 1} \Vert xM\Vert_i$.} of $M$ w.r.t.\  the $\l_i$-norm, $i \in \{1,2,\infty\}$.

 For every $i \in \{1,2,\infty\}$ we have,
\begin{equation}\label{Eq. OP}\tag{6}
\Vert qM - \p \Vert_{i} = \Vert qM - \p M \Vert_{i} \leq \Vert M \Vert_{i} \Vert q - \p \Vert_{i}.
\end{equation}
%As $\Vert M \Vert_{op(2)} = \Vert M \Vert_{op(\infty)}=1$,

It is known that $\Vert M\Vert_\infty$ equals the maximum $\l_1$-norm of a row of $M$ (see,  e.g., Example 5.6.5 on p.345 in \cite{Horn}), and therefore $\Vert M\Vert_\infty=1.$ Also,\footnote{ This follows from the fact that $\Vert M \Vert_2$ equals the maximum singular value of $M$ (see e.g., Example 5.6.6 on p.346 in \cite{Horn}).} $\Vert M \Vert_2 = \Vert M \Vert_\infty=1$. Thus, in view of Eq.\ (\ref{Eq. OP}), any ball (either open or closed) w.r.t.\ to the $\l_2$ or $\l_{\infty}$-norms centered at $\p$ is $M$-absorbing. Moreover, if $M$ is doubly stochastic it is known that\footnote{This follows from the fact that $\Vert M \Vert_{1}$ equals the maximum $\l_1$-norm of a column of $M$ (e.g., Example 5.6.5 on p. 344--345 in \cite{Horn})). } $\Vert M \Vert_1=1$ and therefore in that case any ball (either open or closed) w.r.t.\ the $\l_1$-norm, centered at $\p$, is also $M$-absorbing. See Figure \ref{fugure:absorbing}.

\vskip 1cm

 %{\input{Cav_u-2.tpx}}
\begingroup
\renewcommand\thetheorem{2}
\begin{figure}[htp] \centering{
\includegraphics[scale=0.37]{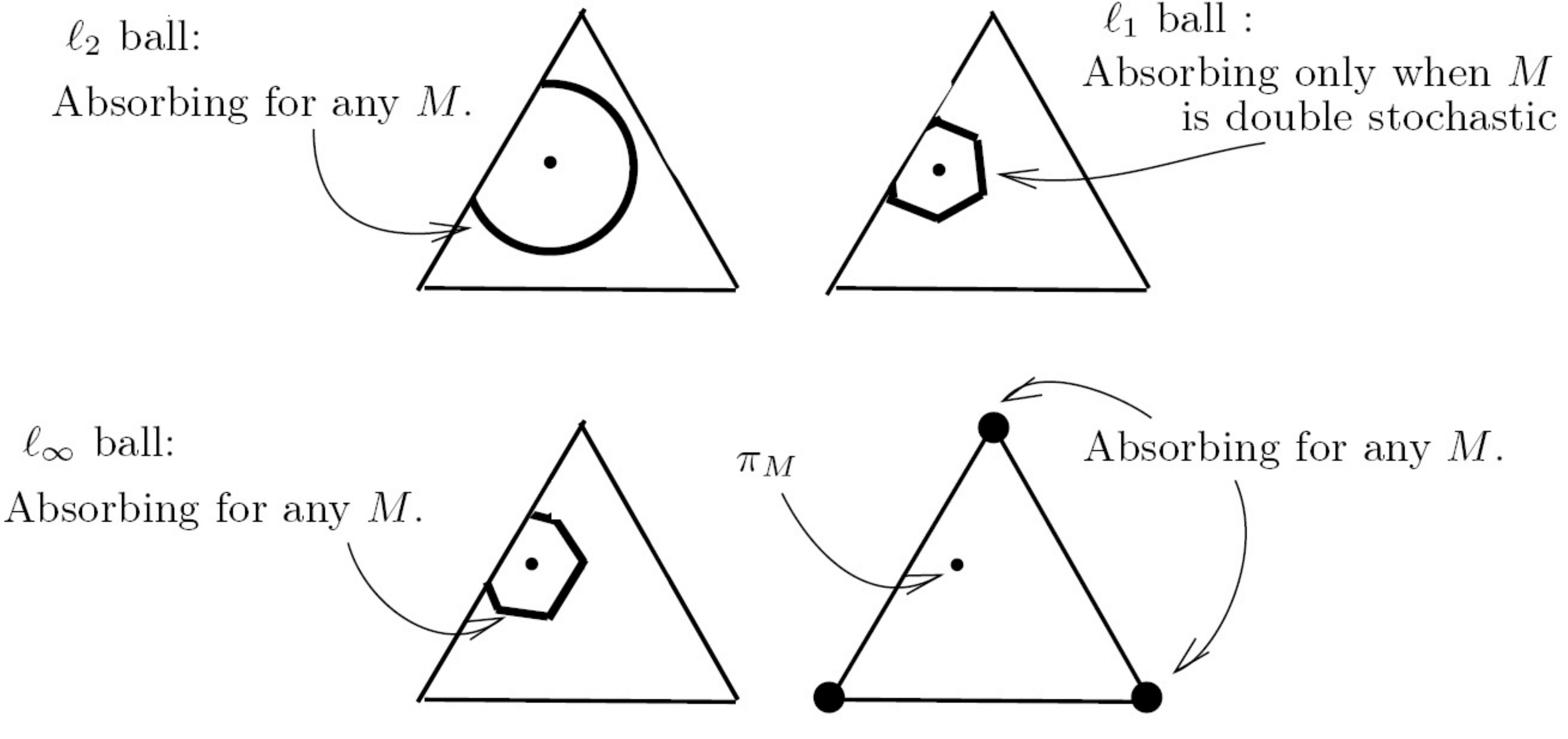}}
\caption{Absorbing sets.}
\label{fugure:absorbing}
\end{figure}
\endgroup

In all the examples above the $M$-absorbing sets contain  $\p$. This is not a coincidence. Indeed,  for every $M$-absorbing set $C$, the image of $\conv(C)$ under the linear map $M$ is also contained in $\conv(C)$. Therefore, by Brouwer's fixed-point theorem,  $M$ possesses a fixed point in\footnote{$\cl \conv(C)$ is the closure of $\conv(C)$.} $\cl \conv(C)$. As the only fixed point of $M$ is $\p$, we deduce that $\p \in \cl \conv(C)$ for every $M$-absorbing set $C$.

We end the discussion on $M$-absorbing sets with the following proposition whose content and proof (delegated to Section \ref{sec:proofs}) may enhance the intuition about absorbing sets.
\begingroup
\renewcommand\thetheorem{1}
\begin{proposition}\label{Prop. Count.}
Let $C$ be an $M$-absorbing set. Then, $C$ contains a countable $M$-absorbing set.
\end{proposition}
\endgroup

\bigskip

\subsection{The Main Theorem.}
\begin{figure}[htp] \centering{
\includegraphics[scale=0.35]{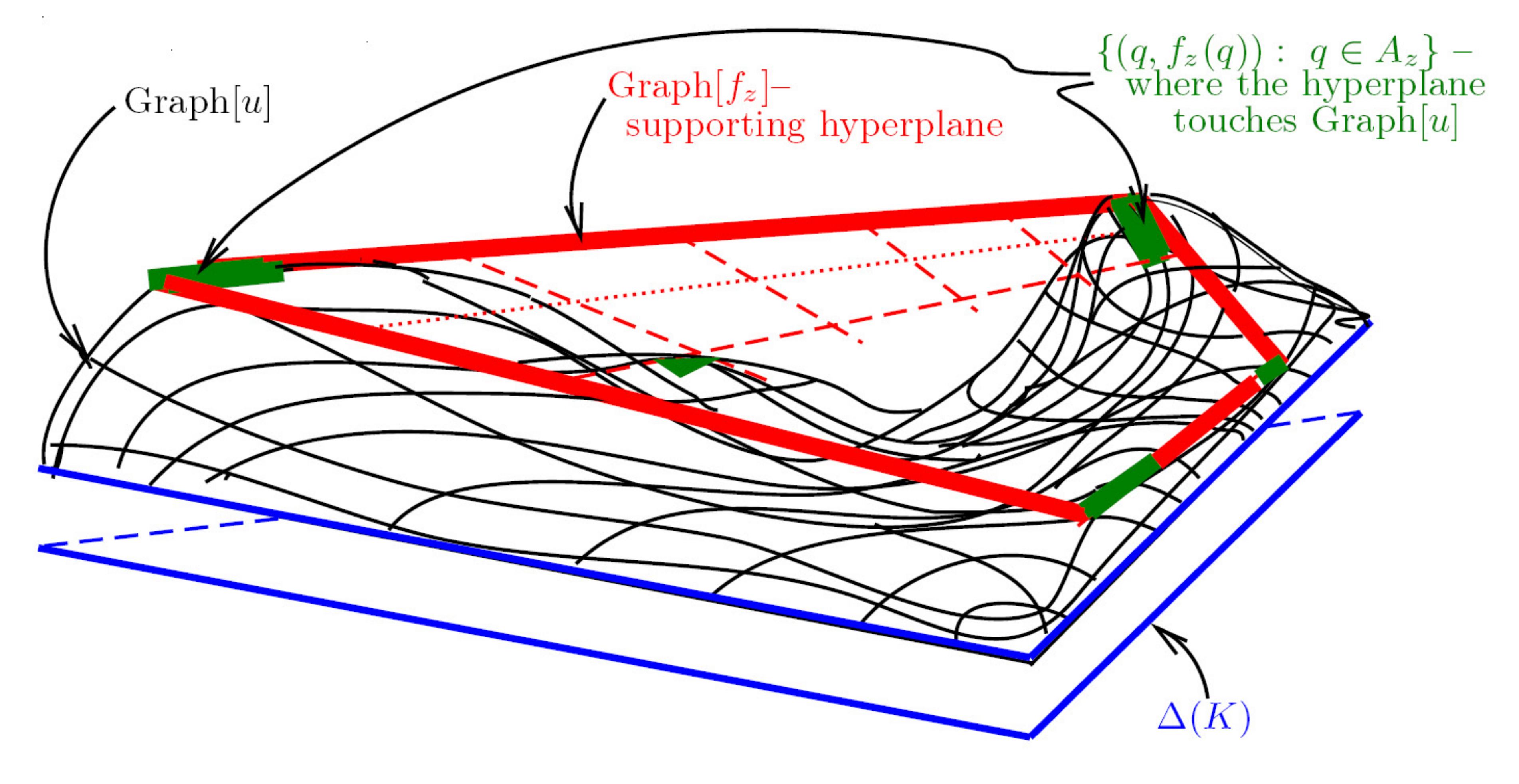}}
\caption{A visualization of $A_z$ for $z \in \Lambda$.}
\label{Figure: A vis of A_z}
\end{figure}

To state our main result we begin with a review of some basic concepts from the theory of concave functions. First, for each $g:\Delta(K) \to \R$ let $\Graph[g] :=\{(q,g(q)):\, q \in \Delta(K)\}$. Since ($\cv$) is a concave function, $\Graph[\cavu]$ can be supported at $(\p,\ccv)$ by a hyperplane. We may parametrize each such supporting hyperplane by a point in $\R^k$ as follows; first, for every $z \in \R^k$ define $f_z:\R^k \to \R$ by $f_z(x):= \ccv + \langle z, q-\p\rangle$. Second, set $$ \Lambda := \{z \in \R^k\,:\, (\cv)(q) \leq f_z(q), \, \forall q \in \Delta(K)\}.$$
As $f_z(\p)= \ccv$ for every $z$, the set $\Lambda$ corresponds to all supporting hyperplanes of $\Graph[(\cv)]$ at $(\p,\ccv)$. For every $z \in \Lambda$ let $$A_z :=\{q \in \Delta(K)\,:\, u(q) = f_z(q)\}.$$ The set $A_z$ can thus be interpreted as the projection to the first $k$ coordinates of the intersection of $\Graph[u]$ with the supporting hyperplane to $\Graph[(\cv)]$ at $(\p,\ccv)$ parametrized by $z$. A visualization of $A_z$ when $k=3$ is given in Figure \ref{Figure: A vis of A_z}.

 %{\input{Cav_u-2.tpx}}

\begingroup
\renewcommand\thetheorem{2}
\begin{proposition}\label{Thm2}
We have the following:
\begin{itemize}
\item[(i)] If $A_z$ contains an $M$-absorbing set for some $z \in \Lambda$, then $v_{\infty} = \Ccv$.
\item[(ii)] If $v_{\infty} = \Ccv$ then for every $z \in \Lambda$, $A_z$ contains a countable $M$-absorbing set.
\end{itemize}
\end{proposition}
\endgroup

Why $M$-absorbing sets contained in the $A_z$'s are of importance? This has to do with the control the sender has on the receiver's beliefs. Indeed, once a belief is in the convex hull of an $M$-absorbing subset $C \subseteq A_z$, $z \in \Lm$, its shift under $M$, which describes the evolution of the posterior in one time period, also lies in $\conv(C)$. At this point in time the sender may send messages that would induce posteriors within $C$, and in particular is $A_z$. As $\cavu$ is an affine function on  $\conv(A_z)$ (see Lemma \ref{Lm Rock} in Section \ref{sec:proofs}), the weighted average of the values of $\cavu$ evaluated at these posteriors, is equal to  the value of $\cavu$ evaluated $\pi_M$.

In the main theorem we summarize the results of Theorem \ref{Thm1} and Proposition \ref{Thm2}.
This theorem characterizes when patient senders can obtain a value close to the maximum possible, the upper bound stated in Theorem \ref{Thm1}.
%\begingroup
%\renewcommand\thetheorem{2}
\begin{theorem}\label{Thm2.5}
The following statements are equivalent:
\begin{itemize}
\item[(i)]For every $\eps>0$ and $p \in \dk$ there exists $0<\d< 1$ such that
\begin{equation}\tag{7}
\vert v_{\lm}(p) - \Ccv \vert < \eps, 	\	\		\	\	\	 \forall \lm > \delta.
\end{equation}
\item[(ii)] There exists  $z \in \Lm$ such that $A_z$ contains an $M$-absorbing set.
\item[(iii)] For every $z \in \Lambda$, $A_z$ contains a countable $M$-absorbing set.
\end{itemize}
\end{theorem}

Note that this characterization is stated in terms of the primitives of the model: $u$ and $M$\ignore{ and $\lm$}. Moreover, it unravels the sensitivity of a patient sender to the interrelationships between $u$ and $M$, as the sets $A_z$, $z \in \Lm$, are fully determined by the former, whereas $M$-absorbing sets are clearly determined by the latter. An interesting question that arises naturally  in this context is how sensitive is a patient sender to such interrelationships. Assume, for instance, that $A_z$ does not contain an $M$-absorbing set for some $z \in \Lm$. Can one quantify the difference between $\ccv$ and $v_{\infty}$ in terms of $u$ and $M$?

\renewcommand\thetheorem{3}
\begin{example}\label{ex: example 2 continued}
\emph{[Example \ref{ex: example 2 continued} continued]} Consider again the matrix $M$ in \emph{Eq.}\ (\emph{\ref{eq: matrix}}) and recall that $\p$ equals $ (0.4,0.6)$. For the function $u$ given in Example \emph{\ref{ex:main}} we have that $A_z$ consisting of the two points $(0,1)$ and $(0.5,0.5)$ in $\dk$ \emph{(}corresponding to the points $0$ and $0.5$ in Figure \emph{\ref{fugure:utility graph}}\emph{)} for every supporting hyperplane $z \in \Lm$. Since this set is not $M$-absorbing, we conclude by  Theorem \emph{\ref{Thm2.5}} that for sufficiently patient sender the value is strictly less than $(\emph{Cav}\, u)(\p)$. However,
if \begin{equation}\label{eq: matrix2}\tag{8}
  M=\begin{pmatrix}
  \ 1/2& 1/2\\
 1/6 & 5/6
\end{pmatrix},
\end{equation}
then $\pi_M=(0.25,0.75)$. Therefore as $A_z =\{(0,1), (0.5,0.5)\}$ identifies with the set of extreme points of the ball of radius $0.25$ w.r.t.\ the $\l_{\infty}$-norm around $\pi_M$, we deduce that $A_z$ is $M$-absorbing for every $z \in \Lm$. Thus, Theorem \emph{\ref{Thm2.5}} ensures that the value $ v_{\lm}(p)$ of a sufficiently patient sender \emph{(}i.e., when  ${\lm}$ is close to $1$\emph{)} is close to $(\emph{Cav}\, u)(\p)=0.512$.

Under both transition matrices, the maximal payoff possible is obtained at $p=0.581$. This point is located in the region where $u$ is equal to $(\Cv)$. We shall prove in Lemma \emph{\ref{lm max}} in Section \emph{\ref{sec:proofs}} that at this point the sender has no incentive to alter the prior belief of the receiver. He therefore reveals no information to the latter. Such a result holds for any continuous $u$ and any point $p$ on which $u$ agrees with $(\Cv)$.
\end{example}

We end the section with a mathematical theorem on $M$-absorbing sets, which is a by product of our main economical result.

\bigskip

\section{Additional Results: A Non-Asymptotic Approach and a Strong Law}\label{sec:non-asymptotic}
\subsection{The value $v_{\lm}$ for every $\lm$.}
\begin{figure}[htp] \centering{
\includegraphics[scale=.4]{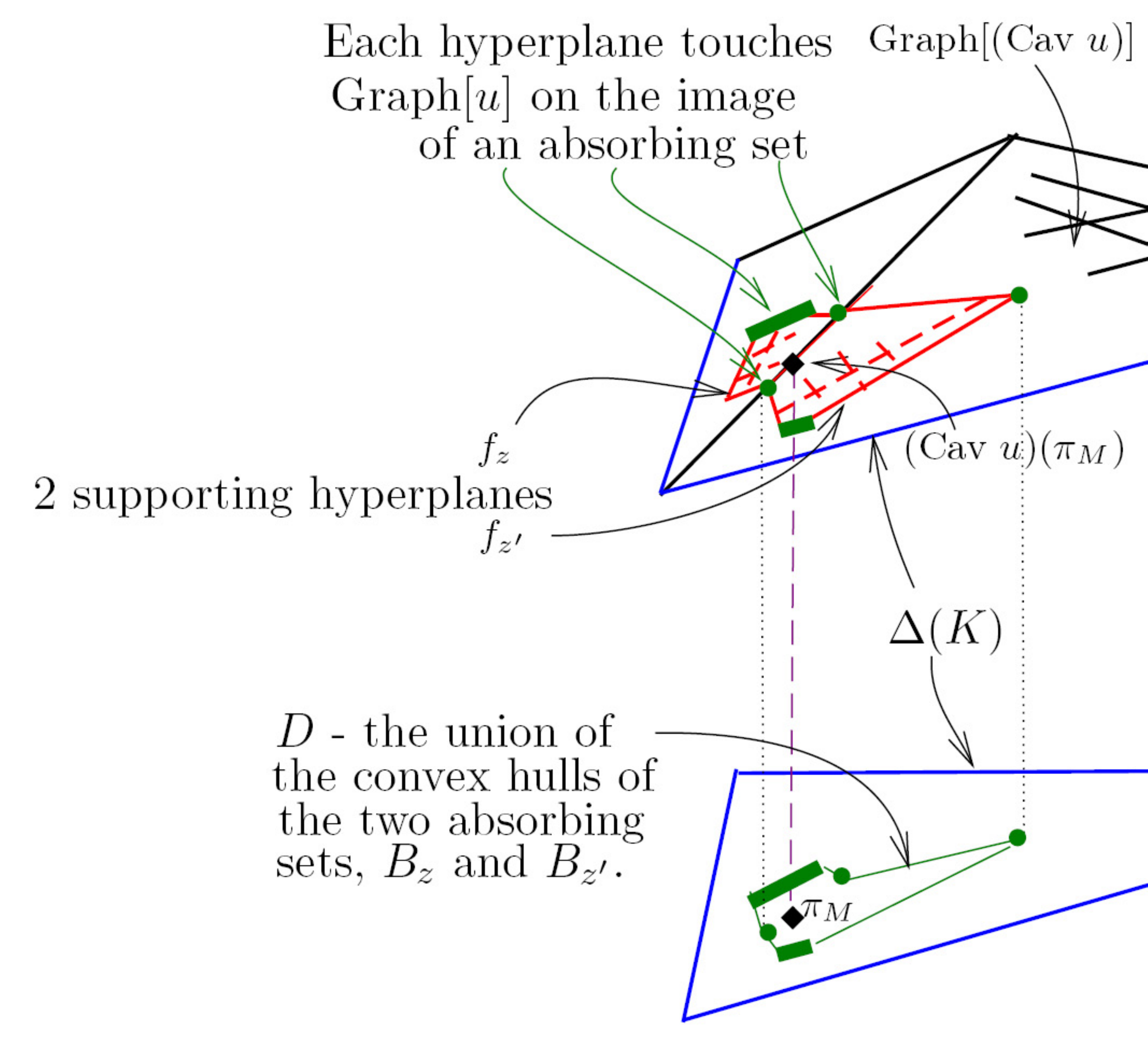}}
%\vskip-.5cm
\caption{A visualization of $D$.}
\label{figure:D}
\end{figure}

As it turns out, the case where $v_{\infty} = \ccv$ encompasses information about the behavior of $v_{\lm}$ across all $\lm \in [0,1)$. To showcase this, we begin by recalling that as any union of $M$-absorbing sets is also $M$-absorbing, by Proposition \ref{Thm2}, if $v_{\infty} = \ccv$, one may associate with each $z \in  \Lambda$ a maximal (w.r.t.\ inclusion) $M$-absorbing set $B_z \subseteq A_z$.   Set $D:= \bigcup_{z \in \Lambda} \conv(B_z)$ (see Figure \ref{figure:D}).

\bigskip

In our next theorem we give an exact formula for $v_{\lm}(p)$ for all $p \in D$ and all $\lm \in [0,1)$ in the case where $v_{\infty}=\ccv$. Therefore, the case $v_{\infty}=\ccv$ is not of a mere asymptotic nature, but rather unravels the full information regarding $v_{\lm}$ for any discount factor $\lm$ on the domain $D \subseteq \dk$. However, outside of $D$, the exact behavior of $v_{\lm}$, even in the case $v_{\infty}=\ccv$, remains an open problem.
\begingroup
\renewcommand\thetheorem{3}
\begin{theorem}\label{Thm3}
Assume that $v_{\infty} = \Ccv$. Then, if $p \in D$ we have
\begin{equation}\label{Eq. E For}\tag{9}
v_{\lm}(p) = (\Cv) \left((1-\lm) p\,  (\emph{Id}_k- \lm M)^{-1}\right), 	\	\ \forall \lm \in [0,1),
\end{equation}
where $\emph{Id}_k$ is the $k\times k$ dimensional identity matrix.
\end{theorem}
\endgroup

Let us now  give some intuition regarding the formula given in Theorem \ref{Thm3}. First,  recall that the sum of an infinite geometric series with common ratio $r \in [0,1)$ and scale factor $1$ equals $1/(1-r)$. As $M$ is stochastic, we will argue in the proof of Theorem \ref{Thm3} that we apply a matrix version of the formula for the sum of infinite geometric series so that $$(1-\lm) p\,  (\text{Id}_k- \lm M)^{-1} = (1-\lm)\sum_{n=1}^{\infty} \lm^{n-1} pM^{n-1}.$$
Moreover, as we shall see in the proof section (e.g., Eq.\ (\ref{M-mrt})), for any signaling strategy $\s$ and prior belief $p \in \dk$ we have that $\E_{p,\s} p_n = pM^{n-1}$. Roughly speaking, the proof of Theorem \ref{Thm3} is divided into two parts. In the first,  we come up with a signaling strategy $\s$ for which the stage payoffs achieve their maximal possible value,  shown to be equal to ($\cv$)$(pM^{n-1})$ for $n \geq 1$. The second part connects $(1-\lm)\sum_{n=1}^{\infty} \lm^{n-1}\,$($\cv$)$(pM^{n-1})$ with the formula given in (\ref{Eq. E For}). Such an argument is valid on $D$, because ($\cv$) is an affine function on each $\conv(B_z)$,  $z \in \Lambda$.
\bigskip

As it turns out, the topological structure of the set $D$ around the point $\p$ may be utilized to derive two-sided bounds on $v_{\lm}$ for every discount factor $\lm \in dk$.
Formally, consider the case where $\p$ is an interior point of $D$ (in the topological space $\dk$ equipped with the $\l_2$-norm). Thus, the Convergence Theorem for Markov Chains (e.g., Theorem 4.9 in \cite{Peres}) guarantees that for every $p \in \dk$ there exists some finite time $N$ so that $pM^N \in D$. By employing bounds on the mixing rates of irreducible and aperiodic Markov chains we obtain the following theorem.
\begingroup
\renewcommand\thetheorem{4}
\begin{theorem}\label{Thm4}
Assume that $v_{\infty} = \Ccv$ and that $\p$ has a full support \emph{(}i.e.,\footnote{int$(D)$ is the interior of $D$.} $\p \in \emph{int}(D)$).  Then, there exists a finite positive integer $N$ such that for every $p \in \dk$ and every $\lm \in [0,1)$ we have the following lower and upper bounds:
\begin{multline}\label{Eq. Thm4}\tag{10}
(1-\lm) \sum_{n=1}^N \lm^{n-1} u (p M^{n-1}) \leq  v_{\lm}(p) - \mathcal{I}_{\lm}(p)\\ \leq (1- \lm) \sum_{n=1}^N \lm^{n-1} (\Cv) (p M^{n-1}),
\end{multline}
where $$\mathcal{I}_{\lm}(p) := \lm^{N} (\Cv)\left((1-\lm) p M^N \,  (\emph{Id}_k- \lm M)^{-1}\right).$$ Moreover, $N$ can be chosen to be of the form $\C \lceil \log_2 r_D^{-1} \rceil$, where $\C>0$ is an integer constant \emph{(}which may depend on $M$\emph{)} and $r_D:= \sup\{r>0:\, B_{\l_1}(\p,r) \subseteq D\}$, where $B_{\l_1}(\p,r)$ is the ball of radius $r$ w.r.t. the $\l_1$-norm centered at $\p$.
\end{theorem}

To illustrate the patience level required to make the bounds in Theorem \ref{Thm4} effective. we provide a numerical example. Assume that $r_D = 2^{-10}$, so that $D$ contains a very small neighborhood of $\p$. Hence,  to attain the value in the left-hand side of Eq.\ (\ref{Eq. Thm4})  the sender can simply avoid sharing any private information along the first $10\,\C$ time periods. As we shall see in the proof of Theorem \ref{Thm4}, after those $10\, \C$ time periods,  the belief of the receiver will enter the region $D$. As soon as this takes place, we will exploit in the proof of Theorem \ref{Thm4} the recursive structure of $\G_{\lm}(p)$ as well as the result of  Theorem \ref{Thm3}, to prescribe to the sender a signaling strategy that will guarantee him  $\mathcal{I}_{\lm}(p)$ in the remaining time periods of the game. Finally, if we choose the discount factor $\lm$ so that $1-\lm^{10\C} = (1-\lm)\sum_{n=1}^{10\C} \lm^{n-1} < \eps$, for some small $\eps$, the lower bound and upper bound on $v_\lm$ in Eq.\ (\ref{Eq. Thm4}) will defer by at most $\eps \times \max_{\dk}|u|$.

\endgroup

\subsection{A Strong Law.}
The next result is concerned with convergence in the strong sense, namely with the $\Pr_{p,\s}$-almost-surely behavior of the random variable $\lim_{\lm \to 1^-}(1-\lm) \sum_{n=1}^{\infty} \lm^{n-1} u(p_n)$ for a specific prior $p$ and a strategy $\s$. The weak-type of convergence employed so far is concerned with the limit of the expectations. In contrast, the next theorem deals with the almost-surely convergence of the payoff, corresponding to a sender interested in the payoffs he actually obtains. These are the payoffs he truly gets along realized paths, not just their expected values.

As it turns out, finite $M$-absorbing sets can be used to deduce a strong law for the distribution of $\lim_{\lm \to 1^-}(1-\lm) \sum_{n=1}^{\infty} \lm^{n-1} u(p_n)$ when  $v_{\infty} = \ccv$.
\begingroup
\renewcommand\thetheorem{5}
\begin{theorem}\label{Thm5}
Assume that $A_z$ contains a finite $M$-absorbing set $C$ for some $z \in \Lm$. Then, there exist an $M$-absorbing subset $Q \subseteq C$ and a strategy $\s \in \Sigma$ such that for every $p \in \emph{conv}(Q)$ it holds that
\begin{equation}\label{Eq.Thm5}\tag{11}
\lim_{\lm \to 1^{-}} (1-\lm) \sum_{n=1}^{\infty} \lm^{n-1} u(p_n) = \Ccv, 	\	\	 \ \Pr_{p,\s}\emph{-}\emph{a.s.}
\end{equation}
Moreover, if $\p \in \emph{int}(Q)$, then  \emph{(\ref{Eq.Thm5})} holds for every $p \in \dk$.
\end{theorem}
\endgroup
In words, under the assumptions of Theorem \ref{Thm5}, for almost every infinite sequences of realizations $(x_1,x_2,...)$ of the Markov chain $(X_n)_{n \geq 1}$, if the sender is patient enough he can guarantee himself by following $\s$ a payoff close to $\ccv$.

\bigskip

\section{When the Main Theorem Holds for every $u$: Homothety.}\label{sec:homothety}
The previous results shed light on the connection between $M$ and $u$. Specifically, the main theorem characterizes when $v_{\infty}=\ccv$ in terms of $M$ and $u$. A natural question arises as to when this result holds for  a fixed $M$ and for every $u$.
To answer this question, we need to introduce the notion of a homothety.

\begin{definition} A linear map, $\psi : \R^k \to \R^k$ is said to be \emph{homothety} with respect to the pair $(v, \b)\in \R^k \times [0,1)$  if $\psi$ maps each point $x \in \R^k$ into the point $\b x + (1-\b) v$. The point $v$ is called the \emph{center} and $\b$ is called the \emph{ratio}.
\end{definition}

It is clear that when $\psi$ is a homothety with respect to $(v, \b)$, the point $v$ is a fixed point of $\psi$. Moreover, $\psi$ reduces the distance from any point $x$ to $v$ by a factor of $\b$. When $v \in \text{int}(\dk)$, or equivalently\footnote{For any measure $\mu$ defined on a finite probability space, we denote by $\text{supp}(\mu)$ its support, i.e., the set of elements  to which $\mu$ assigns a positive probability.} $|\text{supp}(v)|=k$, and $\b \in [0,1)$, the matrix $M^{\psi}$ defined by the homothety $\psi$ is an irreducible aperiodic stochastic matrix. In particular, the stationary distribution of $M$ is $v$. Therefore, we shall say that an irreducible aperiodic stochastic matrix $M$ is a homothety if the mapping $\phi: x \mapsto xM$ is a homothety of $\R^k$ with center $\p$ and ration $\b$, for some $\b \in [0,1)$.

When a stochastic matrix is a homothety, the transition from one state to another follows this law: each state stays unchanged with probability $1-\b$ and moves according to the distribution $v$ to other states with probability $\b$.

We proceed by describing an interesting class of $M$-absorbing sets of a homothety $M$. A set $E \subseteq \dk$ is said to be \emph{star shaped around} $p \in \dk$ if $[p,q] \subseteq E$ for every $q \in E$. In words, assume that an observer is located at the point $p \in \dk$. Then $E$ is star shaped around $p$ if the line of sight, $[p,q]$, to any point $q \in E$ lies entirely in $E$.
Assume now that $E$ is star shaped around $\p$. If $M$ is a homothety, then $qM \in [\p,q) \subseteq E \subseteq \conv(E)$ for every $q \in E$. Hence,  when $M$ is a homothety, every star shaped set around any $\p$ is $M$-absorbing.

Let $M$ be irreducible and aperiodic. Our next result gives a characterization of when $M$ is homothety in terms of $v_{\infty}$. To make the result transparent, note that, by Theorem \ref{Thm1}, $v_{\infty}$ is constant on $\dk$, and as such is simply a function of $u$ and $M$. In our new characterization we let $u$ vary over the space of all continuous functions defined on $\dk$.%, denoted by $C(\dk)$.
Therefore $v_{\infty}$  also varies accordingly.
\begingroup
\renewcommand\thetheorem{6}
\begin{theorem}\label{Thm6}
$M$ is a homothety if and only if $v_{\infty} = \Ccv$ for every continuous function $u$.
\end{theorem}
\endgroup
%%%%%%%%%%%%%%%%%%%

\bigskip

\section{Proofs}\label{sec:proofs}
%%%%%%%%%%%%%%%%%%%
We start this section by reviewing the notion of a \emph{split}, a cornerstone in the field of Bayesian persuasion. This can be described informally as follows: Given a lottery $X$ over $K$ with law $p \in \dk$, to which extent can the sender manipulate (split) $p$ using his signals. The answer, given by Blackwell (1951) and Aumann and Maschler (1995), is that for every choice of distributions $q_1,...,q_{|S|} \in \dk$ and convex weights $(\a_i)_{i=1}^{|S|} \in \Delta(S)$ s.t.\ $\sum_{i=1}^{|S|} \a_i q_i = p$, the sender can correlate his lottery over signals, $Y$, with the lottery $X$, so that on the event that $s^i \in S$ is chosen (having marginal probability $\a_i$) the posterior belief over states becomes $q_i$. This lottery $Y$ will obey the rule
\begin{equation}\tag{12}
\Pr(Y=s^i\,|\, X = \l) = \frac{\a_i q_i^{\l}}{p^{\l}}, \ 	\ \forall i=1,...,|S|, \forall\,\, \l \in K.
\end{equation}
Thus, if we denote by $$\SS_p = \bigg\{\{(q_i,\a_i)\}_{i=1}^{|S|}\,:\, q_i \in \dk\,\, \forall i,\, (\a_i)_{i=1}^{|S|} \in \Delta(S),\, \text{s.t.} \sum_{i=1}^{|S|} \a_i q_i = p \bigg\},$$
the set of all possible splits in $p \in \dk$, then the signaling strategy $\s$ can be interpreted as follows: At the first stage of $\G_{\lm}(p)$, $\s$ specifies which element of $\SS_p$ the sender should pick. Then, for each time period $n \geq 2$, given that $p_{n-1} =q$ for some $q \in \dk$, $\s_n$ prescribes which element of $\SS_{qM}$ the sender should pick. Indeed, prior to getting the signal $s_n$, the receiver updates her belief on the distribution of $X_n$, by taking the multiplication of her posterior belief $q$ on the distribution of $X_{n-1}$ with the transition matrix $M$, resulting in the belief $qM$. It thus follows that the sequence $(p_n)$ satisfies the following important distributional law:
\begin{equation}\label{M-mrt}\tag{13}
\E_{p,\s}\,(p_{n+1}\,|\,p_{n}) = p_{n}M, 	\	\	\	\ \forall n \geq 1.		
\end{equation}
Consequently, $\E_{p,\s}\,p_{n+1} = (\E_{p,\s}p_1) M^{n} = pM^n$ for every $n \geq 1$. In particular if $p=\p$, then $\E_{\p,\s}\,p_{n} = \p$ for every $n \geq 1$.

In light of this fresh view of $\s$, standard arguments (e.g., Theorem 2.20 in \cite{SolanB}) lead to the following `Bellman type' recursive formula for $v_{\lm}(p)$:
\begin{equation}\label{Eq. Bel}\tag{14}
v_{\lm}(p) = \sup_{\{(q_i,\a_i)\}_i \in \SS_p} \bigg\{ (1-\lm)\sum_{i=1}^{|S|} \a_i u (q_i) + \lm \sum_{i=1}^{|S|} \a_i v_{\lm}(q_i M) \bigg\}.
\end{equation}
Consider the operator $\phi:\dk \to \dk$ defined by $\phi(q) = qM$. Since $|S| \geq k$, Carath\'{e}odory's Theorem (see, e.g., Corollary 17.1.5 in \cite{Rock}) implies that the expression on the right-hand side of Eq.\ (\ref{Eq. Bel}) equals $(\text{Cav}\, \{ (1-\lm) u +\lm v_{\lm} \circ \phi\})(p)$. Thus, in conjunction with Eq.\ (\ref{Eq. Bel}) we obtain the following key relation:
\begin{equation}\label{Eq. Rec Cav}\tag{15}
v_{\lm}(p) = (\text{Cav}\, \{ (1-\lm) u +\lm\,(v_{\lm} \circ \phi)\})(p).
\end{equation}
In particular, this shows that the function $v_{\lm}: \dk \to \R$ is concave for every $\lm$. As $\phi$ is linear, $v_{\lm} \circ \phi$ is also concave. Then, by the definition of Cav we infer from Eq.\ (\ref{Eq. Rec Cav}) the following two-sided inequality:
\begin{equation}\label{Eq. 2S}\tag{16}
(1-\lm) \min_{\dk} u + \lm(v_{\lm} \circ \phi) (p) \leq v_{\lm}(p) \leq (1-\lm) \cavu (p) + \lm (v_{\lm} \circ \phi) (p).
\end{equation}
Since the sender can always decide to not reveal any information at $p$, i.e., to choose the split $\{(q_i,\a_i)\}_i \in \SS_p$, where $q_i=p$ for all $i=1,...,|S|$, and thereafter play optimally in the game $\G_{\lm}(pM)$, we also have that $v_{\lm}(p) \geq (1-\lm) u(p) + \lm (v_{\lm} \circ \phi)(p)$. The latter combined with the right hand side of Eq.\ (\ref{Eq. 2S}) gives the following result:
\begingroup
\renewcommand\thetheorem{1}
\begin{lemma}\label{lm max}
Assume that $p\in \dk$ satisfies $u(p) = (\Cv) (p)$. Then, for any $\lm \in [0,1)$, the optimal signaling strategy $\s_{\lm}$ in $\G_{\lm}(p)$, would instruct the sender to not reveal any information at $p$.
\end{lemma}
\endgroup
Set $\vd(p) := \liminf_{\lm \to 1^{-}}v_{\lm}(p)$ and $\vu(p):=\limsup_{\lm \to 1^{-}}v_{\lm}(p)$. Then, by (\ref{Eq. 2S}), $\vd(p) = \vd(pM)$ and $\vu(p)=\vu(pM)$ for every $p \in \dk$. As our Markov chain is irreducible and aperiodic, the Convergence Theorem for Markov chains ensures that $pM^n \to \p$ as $n \to \infty$. A combination of the above two arguments yields the following proposition.
\begingroup
\renewcommand\thetheorem{3}
\begin{proposition}\label{Pr 2}
The functions $\vd$ and $\vu$ are constant on $\dk$.
\end{proposition}
\endgroup
It follows from  Proposition \ref{Pr 2} that on our way to proving Theorem \ref{Thm1} we need to show that $\vd = \vu$. This requires classical tools and techniques from the field of repeated games with incomplete information \cite{Ren. Rev.}. We begin by introducing, for every $N \in \N$ and $p \in \dk$, the $N$-stage game $\G_N (p)$ over the strategy space $\S$ whose payoff is given by the formula
\begin{equation}\tag{17}
\g_N(p,\s) = \E_{p,\s} \left(\frac{1}{N}\sum_{n=1}^N u(p_n)\right).
\end{equation}
The value of $\G_N(p)$ will be denoted by $v_N(p)$. Standard continuity and compactness based arguments (see, e.g., Theorem 2.14 on p.15 in \cite{SolanB}) show that $v_N(p)= \max_{\s \in \S} \g_N(p,\s).$

The following proposition established a number of fundamental properties of $v_N(p)$ which will play an important role in our future proofs.

\begingroup
\renewcommand\thetheorem{4}
\begin{proposition}\label{Pr1}
We have the following:
\begin{itemize}
\item[(i)] $v_N : \dk \to \R$ is concave for every $N \in \N$.
\item[(ii)] For every $N \in \N$, the function $v_N : \dk \to \R$ is Lipschitz \emph{(}w.r.t.\ the $\l^1$-norm\emph{)} with constant $\Vert u \Vert_{\infty}:= \max_{\dk} \vert u \vert$ for every $N \in \N$.
\item[(iii)] The sequence $\{Nv_N(\p)\}_{N}$ is sub-additive.
\item[(iv)]The sequence $\{v_N(\p)\}_{N}$ converges.
\item[(v)] The sequence $\{v_{b^N}(\p)\}_{N}$ is  non-increasing for every $b \in \N$.
\end{itemize}
\end{proposition}
\endgroup
\begin{proof}[Proof of Proposition \emph{\ref{Pr1}}]
The proof of (i) uses the following classical neat argument. Let $q_1,q_2 \in \dk$ and $\a \in (0,1)$ such that $p=\a q_1+(1-\a)q_2$. Prior to the start of $\G_N(p)$ the sender writes a computer program $Z$ which draws either the digit $1$ or $2$ according to a random lottery. This lottery, denoted also $Z$, is dependent of $X_1$, and independent of $(X_n)_{n \geq 2}$. The conditional laws are given as follows:
\begin{equation}\tag{18}
\Pr(Z=1\,|\,X_1 =\l) = 1 - \Pr(Z=2\,|\,X_1 =\l) = \frac{ \a q_1^{\l}}{p^{\l}},	 \	\	 \forall \l \in K.
\end{equation}
Then,  at the first stage the sender sends a messenger\footnote{The messenger cannot be the receiver.} to collect on his behalf the realized value of $X_1$, type it into his program $Z$, and then tell him the output of the program. If the outcome of $Z$ is $1$ ($2$) the sender plays his optimal strategy in $\G(q_1)$ ($G(q_2)$). We now argue that this strategy would guarantee him $\a v_N(q_1)+(1-\a)v_N(q_2)$ in $\G_N(p)$. Indeed, the latter follows by Bayes' law, which implies that the outcome of the program is $1$ (resp., $2$) with probability $\a$ (resp., $1-\a$) and that the posterior distribution of $X_1$ is $q_1$ (resp., $q_2$). The last argument required to depict the situation in which the sender faces $\G_N(q_1)$ (resp., $\G_N(q_2)$) based on the message of the messenger is that after hearing the message ($1$ or $2$), the sender must ask the messenger to inform him of the value of $X_1$ which was realized!.

To prove (ii), observe that by conditioning on the outcome of $X_1$ one has that
\begin{equation}\tag{19}
\g_N(p,\s) = \sum_{\l \in K} p^{\l} \g_N(\delta^{\l},\s),
\end{equation}
for every $\s \in \S$, where $\d^{\l}$ is the Dirac measure supported on $\l \in K$. Therefore, by the triangle inequality, for any $p,q \in \dk$ and any $\s \in \S$,
\begin{equation}\tag{20}
\vert \g_N(p,\s) - \g_N(q,\s) \vert \leq \Vert u \Vert _{\infty} \sum_{\l \in K} \vert p^{\l} - q^{\l} \vert.
\end{equation}
As $\G_N(p)$ can be viewed as a finite game in extensive form, it admits a normal-form game description, and thus (ii) follows from the following basic inequality: for any two zero-sum matrix games $A$ and $B$ of equal dimensions it holds that $\vert \text{val}(A)-\text{val}(B)\vert \leq \Vert A-B\Vert_{\infty}$.

We move on to (iii). For every $\s \in \S$ and every $N,L \in \N$, the $(N+L)$'th stage game payoff $\g_{N+L}(\p,\s)$ equals
\begin{equation}\tag{21}
\frac{N}{N+L}\, \E_{\p,\s} \left[ \frac{1}{N}\sum_{n=1}^N u(p_n)\right] + \frac{L}{N+L}\, \E_{\p,\s} \left[ \frac{1}{L}\sum_{n=N+1}^{N+L} u(p_n)\right].
\end{equation}
As the belief of the receiver at the start of the $(N+1)$'st time period, prior to obtaining the signal $s_{N+1}$, equals $p_N M$, we may bound the latter from above by
\begin{equation}\tag{22}
\frac{N}{N+L}  v_N(\p) + \frac{L}{N+L} \E_{\p,\s} \left[ v_L(p_{N}M)\right].
\end{equation}
By Jensen's inequality applied for the functions $\{v_N\}$, which are known to be concave by (i), in conjunction with the fact that at the initial belief $\p$, all the steps of the sequence $(p_n)$ have expectation $\p$, we obtain that the latter is at most
\begin{equation}\tag{23}
\frac{N}{N+L}  v_N(\p) + \frac{L}{N+L} v_L(\E_{\p,\s}(p_{N}M)) = N v_N(\p)+ L v_L(\p).
\end{equation}
Since this holds for every $\s$, we conclude that$(N+L)v_{N+L} (\p) \leq N v_N(\p)+ L v_L(\p)$, which completes the proof of (iii).

The proof of (iv) can be deduced directly from (iii) based on the following basic result from analysis; if $\{a_n\}$ is a sub-additive sequence, then $\{a_n /n\}$ converges. Lastly, by a repeated use of (iii) we obtain that if $L$ divides $N,$ then
\begin{equation}\tag{24}
Nv_N(\p)\leq (\frac{N}{L}-1) v_{\frac{N}{L}-1}(\p) + L v_L(\p)\leq \cdots \leq \frac{N}{L} (L v_L(\p)).
\end{equation} Therefore, $v_N(\p)\leq v_L(\p)$, which is sufficient to prove (v).
\end{proof}

We are now in position to show that $\vd = \vu$. We split the proof into two steps, the first of which is.
\begingroup
\renewcommand\thetheorem{2}
\begin{lemma}\label{Lm1}
$\lim_{N \to \infty} v_N(\p) \leq \vd$.
\end{lemma}
\endgroup
\begin{proof}[Proof of Lemma \emph{\ref{Lm1}}]
Let $\eps >0$. Define $T_{\eps}:= \min\{n\geq 1:\, qM^n \in B_{\l_1}(\p,\eps), 	\ \, \forall q \in \dk\}$. A well-known bound on the mixing rates of $M$ (see, e.g., Eq.\ (4.34) in Section 4.5 of \cite{Peres}) states that there exists a constant $\C$ (which may depend on $M$) such that $T_{\eps } \leq \C \lceil \log_2 \eps^{-1} \rceil$.

Consider now the following signaling strategy $\s \in \S$. In the first $N_{\eps}$ time periods of the game $\G_{\lm}(\p)$, the sender plays optimally in $\G_{N_{\eps}}(\p)$, where $N_{\eps}$ satisfies (i) $N_{\eps} \geq (\C \lceil \log_2 \eps^{-1} \rceil)/\eps$ and (ii) $v_{N_{\eps}}(\p) \geq \lim_{N \to \infty}v_N(\p)-\eps$. Then, starting from the $(N_{\eps}+1)$'st stage $\s$, asks the sender to ignore all information obtained along the next $T_{\eps}$ time periods, and send the same signal $s_0 \in S$ during these times. Starting from the $(N_{\eps}+T_{\eps}+1)$'st stage, by following $\s$, the sender will play an optimal strategy in $\G_N(p_{N_{\eps}+T_{\eps}} M)$ along the next $N_{\eps}$ time periods, and subsequently play $s_0$ at each of the next $T_{\eps}$ time periods, thereby ignoring all information collected at times $N_{\eps}+T_{\eps}+1,...,2N_{\eps}+2T_{\eps}$.

The extension of the definition of $\s$ to all time periods $n \geq 1$ is done inductively. Let $(b_m)_{m=1}^{\infty}$ be defined by $b_m:= (m-1)N_{\eps}+(m-1)T_{\eps}$. For each $m \geq 1$, the strategy $\s$ instructs the sender to play an optimal strategy in $\G_{N_{\eps}}(p_{b_m} M)$ starting from the $(b_m+1)$'st time period for $N_{\eps}$ periods, followed by sending the fixed signal $s_0 \in S$ at times $b_m+N_{\eps}+1,...,b_{m+1}$.

By the definition of $T_{\eps}$, we have that $p_{b_m} M = p_{b_m - T_{\eps}} M^{T_{\eps}+1} \in B_{\l_1}(\p,\eps)$ for every $m \geq 2$. Hence, by the definition of $\s$, item (ii) of Proposition \ref{Pr1}, and the choice of $N_{\eps}$ we obtain that
\begin{equation}\label{Eq. 1Lm1}\tag{25}
\begin{split}
\frac{1}{N_{\eps}}\sum_{n = b_m+1}^{b_m + N_{\eps}} \E_{\p,\s}(u(p_n)) &= \E_{\p,\s} (v_{N_{\eps}}(p_{b_m}M)) \\
&\geq \E_{\p,\s} (v_{N_{\eps}}(\p) - \Vert u \Vert_{\infty} \eps) \\
&\geq \lim_{N \to \infty} v_{N}(\p) - (\Vert u \Vert_{\infty}+1) \eps,	\	\		\	\	\	\forall m \geq 2.
\end{split}
\end{equation}
As $N_{\eps} / T_{\eps} \geq 1/\eps$, using Eq.\ (\ref{Eq. 1Lm1}) we obtain that \begin{equation}\tag{26}
\liminf_{N \to \infty} \frac{1}{N} \sum_{n=1}^N \E_{\p,\s}(u(p_n)) \geq \lim_{N \to \infty} v_{N}(\p) - O(\eps).
\end{equation}
Combining this with a Tauberian theorem (see, e.g., Theorem 3.1 \ in \cite{SolanB}) and Proposition \ref{Pr 2} we conclude that
\begin{equation}\tag{27}
\vd = \liminf_{\lm \to 1^{-}} v_{\lm}(\p) \geq \liminf_{\lm \to 1^{-}} \g_{\lm}(\p,\s) \geq \lim_{N \to \infty} v_{N}(\p) - O(\eps).
\end{equation}
Since $\eps>0$ was arbitrary all along, the proof is complete.
\end{proof}

The second step towards showing that $\vu = \vd$ is given in the following lemma.
\begingroup
\renewcommand\thetheorem{3}
\begin{lemma}\label{Lm2}
$\vu \leq \lim_{N \to \infty} v_N(\p)$.
\end{lemma}
\endgroup
\begin{proof}[Proof of Lemma \emph{\ref{Lm2}}.]
For each $\lm \in [0,1)$, let $\s_{\lm} \in \S$ be such that\footnote{Such $s_{\lm}$ exists by standard arguments of continuity and compactness, e.g. Theorem 2.14 on p.15 in \cite{SolanB}} $v_{\lm}(\p) = \g_{\lm}(\p,\s_{\lm}).$ Since $u$ is bounded on $\dk$, we may apply the following decomposition (see, e.g., Eq.\ (1) in \cite{Lehrer}) of the discounted average into a convex combination of finite averages:
\begin{equation}\label{lm2Eq1}\tag{28}
\g_{\lm}(\p,\s_{\lm}) = (1-\lm)^2 \sum_{n=0}^{\infty} (n+1)\lm^n \g_{n}(\p,\s_{\lm}).
\end{equation}
Now fix $\eps>0$. By item (iv) of Proposition \ref{Pr1}, there exists $N_0 \in \N$ such that
\begin{equation}\label{lm2Eq2}\tag{29}
\g_{n}(\p,\s_{\lm}) \leq \lim_{N \to \infty} v_N(\p) + \eps,	\	\	\	\forall n > N_0, \,\,\, \forall \lm \in [0,1).
\end{equation}
Next, as $\vert \g_{n}(\p,\s_{\lm}) \vert \leq \Vert u \Vert_{\infty}$ for every $n$ and every $\lm$, we may take $\lm_0 \in [0,1)$ for which
\begin{equation}\label{lm2Eq3}\tag{30}
(1-\lm)^2 \sum_{n=0}^{N_0} (n+1)\lm^n \g_{n}(\p,\s_{\lm}) \leq \eps, 	\	\	\	 \forall \lm > \lm_0.
\end{equation}
Thus, by combining Eqs.\ (\ref{lm2Eq1}), (\ref{lm2Eq2}), and (\ref{lm2Eq3}) we conclude that  $v_{\lm}(\p) \leq  \lim_{N \to \infty} v_N(\p) + 2\eps$ for every $\lm > \lm_0$, thus proving the lemma.
\end{proof}
\bigskip

Combining  Lemmas \ref{Lm1} and \ref{Lm2} we obtain the string of inequalities
\begin{equation*}
\lim_{N \to \infty} v_N(\p) \leq \vd \leq \vu \leq \lim_{N \to \infty} v_N(\p).
\end{equation*}
Hence, $\vu =\vd$. Thus, the proof of Theorem \ref{Thm1} will be concluded if we show the following simple claim to be true.
\begingroup
\renewcommand\thetheorem{1}
\begin{claim}\label{Cl}
For every $\lm \in [0,1)$ and every $N \geq 1$ we have that $v_{\lm}(\p)<\Ccv$ and $v_N(\p)<\Ccv$.
\end{claim}
\endgroup
\begin{proof}[Proof of Claim \emph{\ref{Cl}}.]
The claim follows easily from the fact that for any $\s \in \S$, we have, by Jensen's inequality, that the expected payoff at any time period $n$ satisfies
\begin{equation}\label{Eq. Jen}\tag{31}
\E_{\p,s} u(p_n) \leq \E_{\p,s} (\cv)(p_n) \leq (\cv)(\E_{\p,s}\, p_n) = \ccv.
\end{equation}
\end{proof}

As the first step toward the proof of Proposition \ref{Thm2} and Theorems \ref{Thm3} and \ref{Thm4} we prove the following basic lemma.
\begingroup
\renewcommand\thetheorem{4}
\begin{lemma}\label{Lm Rock}
For every $z \in \Lm$ we have:
\begin{itemize}
\item[(i)]$\Cv (q) = f_z(q)$ for every $q \in \emph{conv}(A_z)$.
\item[(ii)] Let $(\a_i)_{i=1}^m$, $\a_i>0$ for every $i$, $\sum_i \a_i = 1$ and $(q_i)_{i=1}^m \in \dk$ such that $\sum_i \a_iq_i = \p$. If $\sum_i \a_i u(q_i) = \Ccv $, then $q_i \in A_z$ for every $i$.
\end{itemize}
\end{lemma}
\endgroup

\begin{proof}[Proof of Lemma \emph{\ref{Lm Rock}}.]
Let $q \in \conv(A_z)$. Take $(q_i) \in A_z$ and convex weights $(\a_i)$ such that $q = \sum_i \a_i q_i$. Since $\cv$ is concave and $q_i \in A_z$, we have
\begin{equation}\tag{31}
(\cv) (q) \geq \sum_i \a_i (\cv)(q_i) \geq \sum_i \a_i u(q_i) = \sum_i \a_i f_z(q_i) = f_z(q),
\end{equation}
where the last equality follows from the fact that $f_z$ is affine. Since by the definition of $\Lm$ we have $(\cv)(q) \leq f_z (q)$ for every $q \in \dk$, we have shown (i). For (ii) assume that there exists $q_{i_0} \notin A_z$. Then $u(q_{i_0})< f_z(q_{i_0})$, and since $\a_{i_0}>0$ and $z \in \Lm$, we have
\begin{equation}\tag{32}
\ccv = \sum_i \a_i u(q_i) < \sum_i \a_i f_z(q_i) = f_z (\p) = \ccv.
\end{equation}
We reached a contradiction.
\end{proof}

Next, let us now  prove the following proposition, unveiling the special advantages of $M$-absorbing subsets of $A_z$ for $z \in \Lm$.

\begingroup
\renewcommand\thetheorem{5}
\begin{proposition}\label{Pr3}
Assume that for some $z \in \Lm$, the set $A_z$ admits an $M$-absorbing subset $C$. Then,
\begin{equation}\label{Eq. For}\tag{33}
v_{\lm}(p) = (\Cv) \left(\lm p\,  (\emph{Id}_k- \lm M)^{-1}\right), 	\	\ \forall \lm \in [0,1),
\end{equation}
for every $p \in \emph{conv}(C)$, where $\emph{Id}_k$ is the $k$-dimensional identity matrix.
\end{proposition}
\endgroup
\begin{proof}[Proof of Proposition \emph{\ref{Pr3}}.]
To show Eq.\ (\ref{Eq. For}) we will show that a two-sided inequality holds. First, by applying the Jensen's inequality in same fashion as in Eq.\ (\ref{Eq. Jen}), we get that for every $\s \in \S$ and every $\lm \in [0,1)$ it holds that
\begin{multline}\label{Eq. Pr3.1}\tag{34}
\g_{\lm}(p,\s) \leq (1-\lm)\sum_{n=1}^{\infty} \lm^{n-1} (\cv)(\E_{p,\s}\, p_n)\\
 = (1-\lm)\sum_{n=1}^{\infty} \lm^{n-1} (\cv)(pM^{n-1}).
\end{multline}
Since $p \in \conv(C)$ and the set $\conv(C)$ is $M$-absorbing, $pM^{n} \in \conv(C)$ for all $n \geq 1$. Hence, as $\conv(C) \subseteq \conv(A_z)$, item (i) of Lemma \ref{Lm Rock} implies that
\begin{equation}\label{Eq. Pr3.2}\tag{35}
\begin{split}
(1-\lm)\sum_{n=1}^{\infty} \lm^{n-1} (\cv)(pM^{n-1}) &= (1-\lm)\sum_{n=1}^{\infty} \lm^{n-1} f_z(pM^{n-1})	 \\
&= f_z\left((1-\lm)\sum_{n=1}^{\infty} \lm^{n-1}pM^{n-1}\right)\\
&= (\cv) \left( (1-\lm)\sum_{n=1}^{\infty} \lm^{n-1}pM^{n-1}\right)
\end{split}
\end{equation}
where the last equlity follows from item (i) of Lemma \ref{Lm Rock} as well. Moreover, since $M$ is a stochastic matrix, we have the following well-know formula (e.g., Theorem 2.29 in \cite{SolanB}):
\begin{equation}\label{Eq. Pr3.3}\tag{36}
(1-\lm)\sum_{n=1}^{\infty} \lm^{n-1}pM^{n-1} = \lm p\,  (\text{Id}_k-(1-\lm) M)^{-1},	\	 	\ \forall \lm \in [0,1).
\end{equation}
A combination of Eqs. (\ref{Eq. Pr3.1}), (\ref{Eq. Pr3.2}), and (\ref{Eq. Pr3.3}) yields that $v_{\lm}(p)$ is at most $(\cv) \left((1-\lm) p\,  (\text{Id}_k- \lm M)^{-1}\right)$ for every $p \in \conv(C)$ and every $\lm \in [0,1)$. Let us now show that the inverse direction holds as well. We start by defining for each $q \in \dk$ the set $\SS_q^C \subseteq \SS_q$ by
\begin{equation}\label{Eq. Cons. Splt.}\tag{37}
\SS_q^C := \{\{(q_i,\a_i)\}_{i=1}^{|S|}\,:\, q_i \in C \,\,\,\, \forall i=1,...,|S|\}
\end{equation}
Since $|S|\geq k$, Carath\'{e}odory's Theorem shows that $\SS_q^C \neq \emptyset$ whenever $q \in \conv(C)$. Consider now the strategy $\s^C$ defined as follows: at each $n \geq 1$, if $p_{n-1}=q \in \conv(C)$, $\s^C$ will chose an element in $\SS_{qM}^C$; otherwise, if $p_{n-1}=q \in \dk \setminus \conv(C)$, then $\s^C$ will chose some element in $\SS_q$. As $ p \in \conv(C)$, and $\conv(C)$ is $M$-absorbing, we have that under the strategy $\s^C$, $\text{supp}(p_n) \subseteq C$ for every $n \geq 1$. Indeed, we show this by induction on $n$. For $n =1$, since $p \in \conv(C)$, $\text{supp}(p_1) \subseteq C$ by the definition of $\s^z$. Assume now that $\text{supp}(p_n) \subseteq C$ for some $n \geq 1$. Since $C$ is $M$-absorbing, $\text{supp}(p_n M) \subseteq \conv(C)$, and thus by the definition of $\s^C$ we  see that $\text{supp}(p_{n+1}) \subseteq C$ as well. The latter, coupled with $C \subseteq A_z$ implies that the discounted payoff under $\s^C$ can be computed as follows:
\begin{equation}\label{Eq. Pr3.4}\tag{38}
\begin{split}
\g_{\lm}(p,\s^C) & =  (1-\lm)\sum_{n=1}^{\infty}\lm^{n-1} \E_{p,\s^C} [u(p_n)] \\
& = (1-\lm)\sum_{n=1}^{\infty}\lm^{n-1} \E_{p,\s^C}\, [f_z(p_n)]	\\
& =   (1-\lm)\sum_{n=1}^{\infty}\lm^{n-1} f_z\left(\E_{p,\s^C}\, p_n\right)	\\
& =  (1-\lm)\sum_{n=1}^{\infty}\lm^{n-1} f_z\left(pM^{n-1}\right)	\\
& =  (1-\lm)\sum_{n=1}^{\infty} \lm^{n-1} (\cv)(pM^{n-1}),
\end{split}
\end{equation}
where we note that the third equality holds because $f_z$ is affine, and the last equality is a consequence of item (i) of Lemma \ref{Lm Rock}. Hence, by Eqs.\ (\ref{Eq. Pr3.2}) and (\ref{Eq. Pr3.3}) we have that $\g_{\lm}(p,\s^C)$ equals $(\cv) \left((1-\lm) p\,  (\text{Id}_k-\lm M)^{-1}\right)$, thus showing the converse inequality.
\end{proof}

We proceed with the proof of Proposition \ref{Thm2}.
\begin{proof}[Proof of Proposition \emph{\ref{Thm2}}.]
Assume that $C$ is an $M$-absorbing subset of $A_z$ where $z \in \Lm$. Let $q \in \conv(C)$.  First, by Proposition \ref{Pr3} we have $$v_{\lm}(q) = (\cv) \left((1-\lm) q \,  (\text{Id}_k-\lm M)^{-1}\right),	\		\	\forall \lm \in [0,1).$$  Next, since $qM^{n} \to \p$ as $n \to \infty$, we have that $(1/n)\sum_{\l=1}^n qM^{\l-1} \to \p$ as $n \to \infty$. Thus, by a Tauberian Theorem (see, e.g., Theorem 3.1.\ in \cite{SolanB}) we obtain that $(1-\lm)\sum_{n=1}^{\infty} \lm^{n-1}qM^{n-1} \to \p$ as $\lm \to 1^{-}$. Hence, from the identity given in (\ref{Eq. Pr3.3}) we obtain that $(1-\lm) q \,  (\text{Id}_k-\lm M)^{-1}  \to \p$ as $\lm \to 1^{-}$. Moreover, since $\cavu$ is concave on $\dk$ it is also continuous at $\p$ (see, e.g., Theorem 10.1 in \cite{Rock}). A combination of the above arguments with Theorem \ref{Thm1} yields
\begin{multline}\tag{39}
v_{\infty} = \lim_{\lm \to 1^{-}} v_{\lm}(q) =  \lim_{\lm \to 1^{-}} (\cv) \left((1-\lm) q \,  (\text{Id}_k-\lm M)^{-1}\right)\\ = \ccv,
\end{multline}
thus proving item (i) of Proposition \ref{Thm2}.

Let us continue with the proof of item (ii). Assume that $v_{\infty} = \ccv$. We have \emph{(i)} $\lim_{N \to \infty} v_N(\p) = \ccv$, \emph{(ii)} $v_N(\p) \leq \ccv$ for every $N$ (by Claim \ref{Cl}) and \emph{(iii)} $\{v_{b^N}(\p)\}_N$ is non-increasing for every $b \in \N$ (by item (v) of Proposition \ref{Pr1}). A combination of \emph{(i)}, \emph{(ii)}, and \emph{(iii)} shows that $v_N(\p) = \ccv$ for every $N$. Let $\s^N$ be an optimal strategy in $\G_N(\p)$. Denote by $(p_n^N)_n$ the sequence of posteriors induced by $\s^{\lm}$ and the prior probability $\p$. By Jensen's inequality, $\E_{\p,\s^N} [u(p^N_n)] \leq \ccv$ for every $n$. Hence, as $\g_N(\p,\s^N) = \ccv$, we obtain that $\E_{\p,\s^N} [u(p^N_n)] = \ccv$ for every $n=1,...,N$. Fix $\lm \in (0,1)$. We see that
\begin{equation}\tag{40}
\g_{\lm}(\p,\s^N) \geq  (1-\lm)\sum_{n=1}^N \lm^{n-1} \ccv - \lm^N \Vert u \Vert_{\infty}
\end{equation}
for every $N \geq 1$. Letting $N \to \infty$ we get $v_{\lm}(\p) \geq \ccv$. Since by Claim \ref{Cl} the inverse inequality holds as well, we deduce that $v_{\lm}(\p) = \ccv$. Therefore, there exists a strategy $\s^{\lm}$ such that $\g_{\lm}(\p,\s^{\lm}) = \ccv$. Denote the sequence of posteriors induced by $\s^{\lm}$ and the prior probability $\p$ by $(p^{\lm}_n)_n$. By Jensen's inequality, $\E_{\p,\s^{\lm}}\, [u(p^{\lm}_n)] \leq \ccv$, and we therefore obtain that $\E_{\p,\s^{\lm}} [u(p^{\lm}_n)] =  \ccv$ for every $n$. Moreover, as $\text{supp}(p_n^{\lm})$ is finite and $\E_{\p,\s^{\lm}}\, p^{\lm}_n = \p$ for every $n$,  item (ii) of Lemma \ref{Lm Rock} implies that $\text{supp}(p_n^{\lm}) \subseteq A_z$ for every $z \in \Lm$ and every $n$. Set $C:= \bigcup_{n \geq 1} \text{supp}(p_n^{\lm})$. We have $C\subseteq A_z$ for every $z \in \Lm$. Moreover, as the set of signals $S$ of the receiver is finite, we have that  $C$ is the countable union of finite sets, and thus is countable.

We claim that $C$ is $M$-absorbing. Indeed, if $q \in C$, then there exists some $n$ such that $p_n^{\lm} = q$ with positive probability. Since $\E_{\p,\s^{\lm}}(p_{n+1}\,|\,p_n^{\lm} = q) = qM$, we obtain that $qM \in \conv(\text{supp}(p_{n+1}^{\lm})) \subseteq \conv(C)$. To summarize, $C$ is a countable $M$-absorbing subset of $A_z$ for every $z \in \Lm$, as desired.
\end{proof}

We may now deduce the proofs of Theorems \ref{Thm3} and \ref{Thm4} as follows.

\begin{proof}[Proofs of Theorems \emph{\ref{Thm3}} and \emph{\ref{Thm4}}.]
By Proposition \ref{Thm2}, the sets $B_z$, $z \in \Lm$, are well defined (i.e., non-empty). As they are $M$-absorbing, we can apply Proposition \ref{Pr3} to any point $p \in \conv(B_z)$ to get the result of Theorem \ref{Thm3}.

As for the proof of Theorem \ref{Thm4}, let $E:=B_{\l_1}(\p, r)$ where $r<r_D$ satisfies $\lceil \log_2 r^{-1} \rceil = \lceil \log_2 r_D^{-1} \rceil$. By the definition of $r_D$ we have $E \subseteq D$ (see, in the statement of the theorem). By employing a classical bound on the mixing rates of $M$ (e.g., Eq.\ (4.34) in Section 4.5 in \cite{Peres}) we may take $N$ of the form $\C \lceil \log_2 r^{-1} \rceil = \C \lceil \log_2 r_D \rceil$ (where $\C$ is a positive integer which may depend on $M$) so that $pM^N \in E$ for every $p \in \dk$. Thus if the sender does not use his private information along the first $N$ time periods, and then plays optimally starting from the $(N+1)$'st time period, he guarantees $(1-\lm)\sum_{n=1}^{\infty} \lm^{n-1} u(pM^{n-1}) + \lm^N v_{\lm}(pM^N)$. As $pM^N \in D$, Theorem \ref{Thm3} ensures that $\lm^N v_{\lm}(pM^N)=\mathcal{I}_{\lm}(p)$, thus proving the left hand side of Eq.\ (\ref{Eq. Thm4}). On the other hand, by using Jensen inequality for both $\cv$ and $v_{\lm}$, we see that for every $\s \in \S$ and $p \in \dk$ $$\g_{\lm}(p,\s) \leq (1-\lm)\sum_{n=1}^{\infty} \lm^{n-1} \cv(pM^{n-1}) + \lm^N v_{\lm}(pM^N),$$ and thus, since $\lm^N v_{\lm}(pM^N) = \mathcal{I}_{\lm}(p)$ by Theorem \ref{Thm3}, we obtain the right-hand side of Eq.\ (\ref{Eq. Thm4}).
\end{proof}

We move on with the proof of Theorem \ref{Thm5}.
\begin{proof}[Proof of Theorem \emph{\ref{Thm5}}.]
Let $C=\{q_1,...,q_r\}$. Since $C$ is $M$-absorbing, we can assign for each $i=1,...,r$ a distribution $\a^i \in \Delta(C)$ so that $q_i M = \sum_{j=1}^r \a^i_j q_j$. Moreover, by Carath\'{e}odory's Theorem, we can choose $\a^i$ so that $\vert \text{supp}(\a^i) \vert \leq k$ for every $i$. Define the $r\times r$ matrix $W$ by $W_{i,j} := \a^i_j$; $W$ is a stochastic matrix.  Also, define the $r\times k$ dimensional matrix $P$ by $P_{i,\l}:= q_i^{\l}$, where $i \in \{1,...,r\}$ and $\l \in K$. We have the following algebraic relation:
\begin{equation}\label{Eq. Alg1}\tag{41}
P M = W P.
\end{equation}

Let $\Rr$ be a communicating class of $W$ whose states are recurrent (see, e.g., Lemma 1.26 in \cite{Peres}). Denote by $W_{\Rr}$ the restriction of the matrix $W$ to the set of states $\l \in \Rr$. The $|\Rr|\times |\Rr|$  matrix $W_{\Rr}$ is clearly stochastic.   Moreover, since $\Rr$ is a communication class, $W_{\Rr}$ is also irreducible. Next, let us denote by $P_{\Rr}$ the $k \times |\Rr|$  matrix with entries $(P_{\Rr})_{i,\l}:= q_i^{\l}$, where $i \in \Rr$ and $\l \in K$. It follows from Eq.\ (\ref{Eq. Alg1}) that
\begin{equation}\label{Eq. Alg2}\tag{42}
P_{\Rr}M = W_{\Rr} P_{\Rr}.
\end{equation}
As $W_{\Rr}$ is stochastic, Eq.\ (\ref{Eq. Alg2}) implies that the set $Q = \{q_i\}_{i \in \Rr}$ is $M$-absorbing.

Consider the following strategy $\s \in \S$. If $p \in \conv(Q)$, split $p$ into $k$ elements of $Q$ according to some prior $\mu_p \in \Delta(Q)$. Otherwise, ignore private information and send the fixed signal $s_0 \in S$. Assume that $p_{n}=q$ for some $n \geq 1$. If $q \notin \conv(Q)$, the sender ignores his private information and sends the fixed signal $s_0 \in S$. Next, if $q \in \conv(Q)\setminus Q$, $\s$ instructs the sender to choose a split from  $\mathcal{S}_{qM}^Q$ (see (\ref{Eq. Cons. Splt.}) for the definition of $\mathcal{S}_{qM}^Q$). Finally, if $q = q^j \in Q$, $\s$ instructs the sender to split $qM$  into $\{(q_i,w^j_i)\}_{i \in \Rr}$, where $(w^j_1,...,w^j_{|\Rr|})$ is the $j$'th row of $W_{\Rr}$. Note that such a split is available to the sender because each row in $W$ contains at most $k$ non-zero elements, and $k \leq |S|$.

It follows from the definition of $\s$ that for each $p \in \conv(Q)$, the sequence of posteriors $(p_n)$ follows a Markov chain over the state space $Q$ with initial probability $\mu_p$ and transition rule given by the stochastic matrix $W_{\Rr}$. Since the latter is irreducible, we may employ the Ergodic Theorem for Markov chains (e.g., Theorem C.1 in \cite{Peres}) to obtain that
\begin{equation}\tag{43}
\lim_{N \to \infty} \frac{1}{N}\sum_{n=1}^N u(p_n) = \sum_{i=1}^{|\Rr|} \nu_i u(q_i),		\	\	 \ \Pr_{p,\s}\emph{-}\text{a.s.}
\end{equation}
where $\nu = (\nu_i)_{i=1}^{|\Rr|}$ is the unique stationary distribution of $W_{\Rr}$.
Furthermore, since $Q \subseteq A_z$, we have that
\begin{equation}\tag{44}
\sum_{i=1}^{|\Rr|} \nu_i u(q_i)  =  \sum_{i=1}^{|\Rr|} \nu_i f_z(q_i)
 =  f_z\left(\sum_{i=1}^{|\Rr|} \nu_i q_i \right)
=  (\cv) (\sum_{i=1}^{|\Rr|} \nu_i q_i),
\end{equation}
where the last equality follows from item (i) of Lemma \ref{Lm Rock}. Next, by multiplying by $\nu$ from the left  both sides of Eq.\ (\ref{Eq. Alg2})  we obtain
\begin{equation}\tag{45}
\nu P_{\Rr}M =  \nu W_{\Rr} P_{\Rr} = \nu P_{\Rr},
\end{equation}
which together with the uniqueness of $\p$ implies that $\nu P_{\Rr} = \p$. However, as $\p = \nu P_{\Rr} = \sum_{i=1}^{|\Rr|} \nu_i q_i$, we deduce that
\begin{equation}\label{Eq. Erg. L}\tag{46}
\lim_{N \to \infty} \frac{1}{N}\sum_{n=1}^N u(p_n) = \ccv, 	\	\		\	\Pr_{p,\s}\emph{-}\text{a.s.}
\end{equation}
Therefore, by a Tauberian theorem  (see, e.g., Theorem 3.1.\ in \cite{SolanB}),
\begin{equation}\label{Eq. T-E.}\tag{47}
\lim_{\lm \to 1^{-}} (1-\lm)\sum_{n=1}^{\infty} \lm^{n-1} u(p_n) = \ccv,		\	\	 \ \Pr_{p,\s}\emph{-}\text{a.s.}
\end{equation}
Finally, assume that $\p \in \text{int}(\conv(Q))$. Then, since $M$ is irreducible and aperiodic, for some finite time period $N$ for every $p \in \dk$. Then, by the definition of $\s$, we see that (\ref{Eq. Erg. L}) holds for every $p \in \dk$ and thus so does (\ref{Eq. T-E.})
\end{proof}

The proof of Proposition \emph{\ref{Prop. Count.}} may enhance the intuition about absorbing sets.
\begin{proof}[Proof of Proposition \emph{\ref{Prop. Count.}}]
By Carath\'{e}odory's Theorem (see, e.g., Corollary 17.1.5 in \cite{Rock}) we may assign with each $q \in C$ $k$ distributions $w_1(q),...,w_k(q) \in C$ such that $qM \in \conv(\{w_1(q),...,w_k(q)\}$. Define a correspondence $\xi: C \to 2^C$ by $\xi(q) = \{w_1(q),...,w_n(q)\}$. In particular,  $qM \in \xi(q)$. The set $\mathcal{A}(q) := \bigcup_{n=1^{\infty}} \xi^{n-1}(q)$ is $M$-absorbing for  every $q \in C$, where $\xi^{n-1}$ is the composition of $\xi$ with itself $n-1$ times. Indeed, let $w \in \mathcal{A}(q)$, and let $n \geq 1$ such that $w \in \xi^{n-1}(q)$. By the definition of $\xi$ we have that $wM \in \conv(\xi(w)) \subseteq \conv(\xi^{n}(q)) \subseteq \conv(\mathcal{A}(q))$ as desired.
\end{proof}

We end the proof section with the proof of Theorem \ref{Thm6}.
\begin{proof}[Proof of Theorem \emph{\ref{Thm6}}.]
Suppose that $M$ is a homothety and let us fix $u \in C(\dk)$. As $u$ is continuous, Carath\'{e}odory's Theorem (see, e.g., Corollary 17.1.5 in \cite{Rock}) implies that there exist $q_1,...,q_m \in \dk$, $m \leq k$,  and positive convex weights $(\a_i)_{i=1}^m$ such that $\p = \sum_{i=1}^m \a_i q_i$ and $\ccv = \sum_{i=1}^{m} \a_i u(q_i)$. Hence, by item (ii) of Lemma \ref{Lm Rock}, $q_i \in A_z$ for every $i$ and every $z \in \Lambda$. Therefore, $\p \in \conv(A_z)$ for every $z \in \Lambda$. The latter coupled with the convexity of $\conv(A_z)$ implies that $\conv(A_z)$ is star shaped around $\p$. Hence, since $M$ is a homothety we get that $\conv(A_z)$ is $M$-absorbing for any $z \in \Lambda$. By the definition of an $M$-absorbing set, $A_z$ must also be $M$-absorbing for any $z \in \Lambda$. By Proposition \ref{Thm2}, we deduce that $v_{\infty} = \ccv$, proving the first direction of Theorem \ref{Thm6}.

Suppose now that $v_{\infty}=\ccv$ for every $u \in C(\dk)$. For each $i \in K$, let $e_i \in \dk$ be the Dirac measure concentrated on the $i$'th coordinate of $\R^k$. Fix  $i \in K$ and consider for each $n \geq 1$ the vector $e_i^n = \p + (\p-e_i)/n$. Clearly, $\p \in [e_i,e_i^n]$ for all $n$. Next, as $\p \in \text{int}(\dk)$, there exists $N_i$ such that $e_i^n \in \dk$ for every $n \geq N_i$. For each $n \geq N_i$ we take an element $u_{i,n} \in C(\dk)$ satisfying $u_{i,n}(q) = 1 $ for $q \in \{e_i,e_i^n\}$ and $u_{i,n}(q) < 1$ for $q \in \dk \setminus \{e_i,e_i^n\}$. We have that $\ccv = 1$. Moreover, by the definition of $u_{i,n}$, $0 \in \Lambda$ (for $\Lambda$ corresponding to $u=u_{i,n}$), because the hyperplane $f_0 (x) = \ccv = 1$ for all $x \in \R^k$, supports $(\cv)$ at $\p$. As $A_0 = \{e_i, e_i^n\}$, Proposition \ref{Thm2} shows that $\{e_i, e_i^n\}$ contains an $M$-absorbing subset. However, as $M$ has a unique stationary distribution  $\p \notin \{e_i, e_i^n\}$  we get that neither $\{e_i\}$ nor $\{e_i^n\}$ is $M$-absorbing. Therefore, $\{e_i, e_i^n\}$ must be $M$-absorbing for every $n \geq N_i$. In particular, $e_i M \in (e_i,e_i^n]$ for every $n \geq N_i$. Since $e_i^n \to \p$ as $n \to \infty$, we obtain that $e_i M \in (e_i,\p]$. As $i$ was arbitrary, we have shown that for each $i \in K$ there exists $\b_i \in [0,1)$ such that $e_i M = \b_i e_i + (1-\b_i)\p$. Since $q \to Mq$ is a linear operator, to prove that $M$ is a homothety is suffices to show that $\b_i = \b_j$ for all $i \neq j \in K$.

The proof now bifurcates according to the dimension of $\dk$. First, let us assume that $k =2$. Since $M$ is irreducible, there exists a unique $\a \in (0,1)$ so that $\p = \a e_1 + (1-\a)e_2$. We have
\begin{equation}\label{Eq.1 Thm6}\tag{48}
\begin{split}
\p M & =  (\a e_1 + (1-\a)e_2 )M  \\
& = \a (e_1 M) + (1-\a)(e_2 M) \\
& =  \a (\b_1 e_1 +(1-\b_1)\p)+ (1-\a)(\b_2 e_2 + (1-\b_2)\p).
\end{split}
\end{equation}
By plugging $\p = \a e_q + (1-\a)e_2$ into the last expression in Eq.\ (\ref{Eq.1 Thm6}) and using simple algebraic manipulations, we get that the convex weight $\rho$ of $e_1$ in the convex decomposition of $\p M$ to $e_1$ and $e_2$ equals
$$\a \b_1 + \a^2(1-\b_1) + (1-\a)(1-\b_2)\a.$$ However, as $\p = \p M$, we must have that $\rho =\a$. After some further simple algebraic manipulations, one gets that the equality $\rho = \a$ is equivalent to  $\b_1 - \b_2 = \a (\b_1 - \b_2)$. As $\a \in (0,1)$,  we obtain that $\b_1 = \b_2$, thus proving that $M$ is a homothety whenever $k=2$.

Next, let $k \geq 3$. Assume that $\b_i < \b_j$ for some $i \neq j \in K$. Define $v = (e_i +e_j)/2$. Since $\vert \text{supp}(v) \vert =2$, whereas $ \vert \text{supp}(\p) \vert =k \geq 3$, we have that $\p \neq v$. Consider for each $n \in \N$ the vector $v_n = \p + (\p-v)/n$. Then $\p \in [v,v_n]$ for every $n$. Moreover, since $\p \in \text{int}(\dk)$ there exists $N_0$ such that for every $n \geq N_0$ we have $v_n \in \dk$. As at the beginning of the proof we take for each $n \geq N_0$ an element $u_{n} \in C(\dk)$ satisfying $u_{n}(q) = 1 $ for $q \in \{v,v_n\}$ and $u_{n}(q) < 1$ for $q \in \dk \setminus \{v,v_n\}$. Hence, by arguing as before for $e_i, e_i^n$ and $u_i^n$, only this time for $v,v_n$ and $u_n$,  we obtain that $v M \in (v,v_n]$ for every $n \geq N_0$. As $v_n \to \p$  as $n \to \infty $ we see  that $v M \in (v,\p]$. On the other hand,
\begin{equation}\label{Eq.2 Thm6}\tag{49}
\begin{split}
v M & =  \frac{1}{2}\, (e_i M + e_j M) \\
& =  \frac{1}{2}\, (\b_i e_i + (1-\b_i)\p) + \frac{1}{2}\,(\b_j e_j + (1-\b_j)\p)\\
& =  \left(1 - \frac{\b_i}{2} - \frac{\b_j}{2}\right)\p + \b_i v + \frac{1}{2}\, (\b_j - \b_i)e_j.
\end{split}
\end{equation}
As $0 \leq \b_i < \b_j<1$, this implies that $v M$ lies in the relative interior of the triangle defined by $\p$, $v$, and $e_j$. This of course contradicts  the fact that $v M \in (v,\p]$. Hence, $\b_i = \b_j$ for every $i\neq j \in K$, thus proving that $M$ is a homothety.
\end{proof}

\end{document}